\documentclass[aps,twocolumn,secnumarabic,groupedaddress,floatfix,amsfonts]{revtex4}
\usepackage[dvips]{graphicx,color}
\begin{document}
\title{Experimental consistency in parton distribution fitting}
\author{Jon Pumplin}
\affiliation{
Michigan State University, East Lansing MI 48824, USA 
}
\date{\today}
\preprint{MSUHEP-090831}
\begin{abstract} 
The recently developed ``Data Set Diagonalization'' method (DSD) is applied to 
measure compatibility of the data sets that are used to determine parton 
distribution functions (PDFs).  Discrepancies among the experiments are 
found to be somewhat larger than is predicted by propagating the published 
experimental errors according to Gaussian statistics.  The results support 
a tolerance criterion of $\Delta\chi^2 \approx 10$ to estimate the 
90\% confidence range for PDF uncertainties.  No basis is found in the data 
sets for the much larger $\Delta\chi^2$ values that are in current use; though
it will be necessary to retain those larger values until improved methods can 
be developed to take account of systematic errors in applying the theory.
The DSD method also measures how much influence each experiment has on 
the global fit, and identifies experiments that show significant tension 
with respect to the others.  The method is used to explore the contribution 
from muon scattering experiments, which are found to exhibit the largest 
discrepancies in the current fit.
\end{abstract} 
\maketitle
\section{Introduction
\label{sec:introduction}}
Interactions at high energy colliders such as the Tevatron and LHC are 
interpreted according to Quantum 
Chromodynamics (QCD) and Electroweak theory on the basis of collisions between 
partons.  The analysis of collider data therefore relies on knowing the parton 
distribution functions (PDFs) that describe probability densities for the gluon 
($g$) and quark partons ($u$, $d$, $c$, $s$, $b$) and their antiquarks 
($\bar{u}$, $\bar{d}$, $\bar{c}$, $\bar{s}$, $\bar{b}$) in the proton, as a 
function of momentum fraction $x$ and QCD factorization scale $\mu$.  
In keeping with their importance, there is a sizeable industry in attempting 
to determine the PDFs \cite{cteq60,cteq66,CT09,MSTW08,NNPDF}.

An integral part of the PDF effort is the need to estimate the uncertainty of 
the results.  An obvious component of that uncertainty comes from the reported 
errors in the data.  It has become standard practice \cite{cteq66,CT09,MSTW08} 
to inflate the uncertainties obtained in this way, motivated in part by a 
notion that disagreements between different experiments in the global fit 
signal the presence of unknown systematic errors in the experiments---or else 
they indicate important systematic errors in the theory, which could for 
example be introduced by the perturbative approximations to QCD.

The recently-invented method of Data Set Diagonalization (DSD) \cite{DSD} 
offers a direct assessment of the contribution from each experiment to the 
global fit, and provides a statistical measure of the consistency between 
each experiment and the others.  The method was illustrated in \cite{DSD} 
by applying it to three of the experiments in a contemporary PDF 
analysis \cite{CT09}. That study is extended in this paper to systematically 
examine the contribution and consistency of every experiment in the analysis.

The DSD study is important for two reasons.  First, the overall level of 
consistency among the experiments provides quantitative information on how to 
assign uncertainty estimates to predictions based on the global fit.  Second, 
the study identifies experiments whose implications are in disagreement 
with the consensus of the others, due to unknown theoretical or 
experimental problems.

\section{The PDF fitting paradigm
\label{sec:paradigm}}
In current practice, one attempts to determine 
$u(x)$, $\bar{u}(x)$,
$d(x)$, $\bar{d}(x)$,
$s(x)$, $\bar{s}(x)$, $g(x)$ at some low QCD scale $\mu_0$.
The distributions at all higher scales are then 
given by the QCD renormalization group DGLAP 
equations.  The $c$ and $b$ distributions are generally assumed to arise only 
from this perturbatively calculable evolution in $\mu$; and available 
data are consistent with $s(x) = \bar{s}(x)$. This leaves 6 
unknown functions of $x$ to be determined from experiment.  These functions
are further constrained theoretically only by the number sum rules 
$\int_0^1 \, (u(x) - \bar{u}(x)) \, dx = 2$ and  
 $\int_0^1 \, (d(x) - \bar{d}(x)) \, dx = 1$,
the momentum sum rule, 
some theoretical predictions on limiting behavior at $x \to 0$ and $x \to 1$,
positivity, and notions of expected smoothness.

In the paradigm used here, the parton distributions at $\mu_0$ are expressed 
as functional forms in $x$, with a large number of adjustable parameters.
The parameter values  are determined by a ``global analysis'' 
in which data from a wide variety of experiments are fitted simultaneously.  
No single experiment directly measures any one the basic distributions; but 
the workings of QCD tie each data point to a different convolution integral 
over the distributions, and hence to a different combination of the unknown 
parameters.  

\section{The DSD method
\label{sec:DSDmethod}}
This Section summarizes the DSD method \cite{DSD}.  As a further aid to 
understanding it, the method is illustrated in the Appendix by a simple 
explicit example.

The $\chi^2$ measure of the quality of fit to the full body of data is a 
function of parameters $a_1$,\dots,$\, a_n$ that define the parton 
distributions at scale $\mu_0$ (here $n=24$ and 
$\mu_0 = 1.4 \, \mathrm{GeV}$).  
The best-fit PDF set is found by minimizing $\chi^2$ with respect to those 
parameters.  The uncertainty range of the fit is estimated as the region in 
parameter space that is sufficiently close to this 
minimum: $\chi^2 \, \le \, \chi_{\mathrm{min}}^2 \, + \, \Delta\chi^2 \,$.

The dependence of $\chi^2$ on $\{a_i\}$ can be expanded about the minimum 
through second order using Taylor series.  The eigenvectors of the quadratic 
form that governs that expansion can be used as basis vectors to obtain a 
linear transformation to new coordinates for which
\begin{equation}
  \chi^2 \, =  \, \chi_{\mathrm{min}}^{\,2} \, + \, 
               \sum_{i=1}^n y_i^{\, 2} \; . 
  \label{eq:diagchi} 
\end{equation}
This is known as the Hessian method \cite{Hessian}.  The DSD method \cite{DSD} 
builds upon it by simultaneously diagonalizing $\chi^2$ and the contribution 
to $\chi^2$ from a subset of the data, such as a single one of the 
experiments.  This is done as follows. Let $\chi_{\mathbf{S}}^{\, 2}$ be the 
contribution to $\chi^2$ from the subset.  In the neighborhood of the global 
minimum, $\chi_{\mathbf{S}}^{\, 2}$ can be expanded through second order in 
the coordinates $\{y_i\}$ by again using Taylor series.  The eigenvectors of 
the second-derivative matrix which appears in that expansion provide a further 
linear transformation which diagonalizes its quadratic form, without spoiling 
Eq.~(\ref{eq:diagchi}).  Combining the two linear transformations yields a 
single linear transformation of the fitting parameters $\{a_i\}$ to new 
parameters $\{z_i\}$ for which
\begin{eqnarray}
  \chi^{\, 2} &=&  \chi_{\mathbf{S}}^{\, 2} \, + \,
                   \chi_{\mathbf{\overline{S}}}^{\, 2} \, = \,  
                   \sum_{i=1}^n z_i^{\, 2} 
   \label{eq:GammaAC1} \\
  \chi_{\mathbf{S}}^{\, 2}  &=&   
  \sum_{i=1}^n \gamma_i \, (z_i - A_i)^2  \, + \, \mathrm{const} 
   \label{eq:GammaAC2} \\
  \chi_{\mathbf{\overline{S}}}^{\, 2} &=&  
  \sum_{i=1}^n \, (1 - \gamma_i)\, (z_i - C_i)^2 \, + \, \mathrm{const} \; ,
   \label{eq:GammaAC3}
\end{eqnarray}
where $\, \gamma_i A_i \, + \, (1-\gamma_i) C_i \, = \, 0$.  Assuming 
that $0 < \gamma_i < 1$, Eqs.~(\ref{eq:GammaAC2}--\ref{eq:GammaAC3}) can be 
written in the form
\begin{eqnarray}
  \chi_{\mathbf{S}}^{\, 2}  &=&   
   \sum_{i=1}^n \left(\frac{z_i - A_i}{B_i}\right)^2 \, + \, \mathrm{const} \nonumber \\
  \chi_{\mathbf{\overline{S}}}^{\, 2} &=&  
    \sum_{i=1}^n \left(\frac{z_i - C_i}{D_i}\right)^2 \, + \, \mathrm{const} \; ,
   \label{eq:BetaGamma}
\end{eqnarray}
which cries out to be be interpreted as independent measurements:
\begin{eqnarray}
  \mbox{$\mathbf{S}$:}\;\;  z_i &=& A_i \, \pm \, B_i \nonumber \\
  \mbox{$\mathbf{\overline{S}}$:}\;\;  z_i &=& C_i \, \pm \, D_i \; .
   \label{eq:CentralResult1}
\end{eqnarray}
The parameters $\gamma_i$ determine the precision of these 
measurements through
\begin{eqnarray}
     B_i &=& 1/\sqrt{\gamma_i}   \nonumber \\
     D_i &=& 1/\sqrt{1-\gamma_i} \; .
   \label{eq:CentralResult2}
\end{eqnarray}

In the PDF analysis, the largest values of $\gamma_i$ that appear are 
$\, \sim \! 0.9\,$.  Most of the $\gamma_i$ are smaller than that, since
most properties of the 
global fit are significantly constrained by more than one experiment---both 
because different kinds of experiments are strongly linked by QCD, and because 
many of the key measurements have been made more than once (often by more 
than one experimental group).
The study in Sec.\ \ref{sec:Results} reports all of the results with 
$\gamma_i \ge 0.1$.  Directions for which $\gamma_i < 0.1$ can be neglected, 
since for these directions, the uncertainty of $\mathbf{S}$ is at least 3 times 
larger than the uncertainty from the other experiments, so it contributes little 
to the weighted average.  In practice 
some $\gamma_i$ even come out negative.  When that happens, it indicates that 
$\mathbf{S}$ is so insensitive to $z_i$ in the allowed range 
$|z_i| \lesssim 1$ that the quadratic approximation has broken down for that 
experiment along that direction.  Since $\mathbf{S}$ is insensitive to $z_i$ 
along such directions, it is correct to ignore them along with the other 
directions for which $\gamma_i < 0.1\,$.

The new coordinates are chosen such that the average of the two measurements 
(\ref{eq:CentralResult1}), weighted by their uncertainties, gives
\begin{equation}
   z_i = 0 \, \pm \, 1
\end{equation}
according to Eq.\ (\ref{eq:GammaAC1}).  The difference between the two 
measurements 
(\ref{eq:CentralResult1}) 
provides a direct measure of the consistency between $\mathbf{S}$ and its
complement $\mathbf{\overline{S}}$.  That difference can be 
expressed in standard deviations as
\begin{equation}
   \sigma_i \, = \, \frac{|A_i - C_i|}{\sqrt{B_i^{\, 2} + D_i^{\, 2}}} \, = \,
   \sqrt{\gamma_i \, (1 - \gamma_i)} \, |A_i - C_i| \; .
\end{equation}
 
The parameter $\gamma_i$ characterizes the importance of experiment 
$\mathbf{S}$, while the parameter $\sigma_i$ characterizes its 
consistency with $\mathbf{\overline{S}}$, along direction $z_i$.  
In the next Section, these key parameters are evaluated for every 
experiment in the PDF global fit.

\section{Results from the DSD method
\label{sec:Results}}

We study a body of input data that is nearly the same as was used in the recent 
CT09 analysis \cite{CT09}.  
The parametrization of the PDFs is identical to CT09, with the same 
24 free parameters.   The definition of $\chi^2$ used here is 
just the sum over data points of ((data-theory)/error)$^2$, 
except for including correlated systematic experimental errors for all data 
sets for which these have been published.  Unlike in CT09, no weight factors or 
penalties are applied in $\chi^2$ to emphasize particular experiments.

\begingroup
\begin{table*}[htb]
  \begin{center}
\renewcommand\arraystretch{0.90}
\null
\begin{tabular}{||c|c|r|r||l||}
\hline
Process & Expt & N $\,$ &  $\sum_i \gamma_i$ & 
$(\gamma_1,\, \sigma_1), \, (\gamma_2,\, \sigma_2),\, \dots$ \\
\hline
\hline
$ e^+ \, p \to e^+ \, X$ & H1 NC \hfill \cite{Adloff:2000qk} &$\,$115$\,$& 2.10 &  
$(0.72,\,0.01)$ $(0.59,\,\mathbf{3.02})$ $(0.43,\,0.20)$ $(0.36,\,1.37)$  \\
$ e^- \, p \to e^- \, X$ & H1 NC \hfill \cite{Adloff:2000qj} & 126$\,$ & 0.30 &  
$(0.30,\,0.02)$  \\
$ e^+ \, p \to e^+ \, X$ & H1 NC \hfill \cite{Adloff:2003uh} & 147$\,$ & 0.37 &  
$(0.21,\,0.06)$ $(0.16,\,0.83)$  \\
$ e^+ \, p \to e^+ \, X$ & H1 CC \hfill \cite{Adloff:1999ah} & 25$\,$ & 0.24 &  
$(0.24,\,0.00)$ \\
$ e^- \, p \to \nu \, X$ & H1 CC \hfill \cite{Adloff:2000qj} & 28$\,$ & 0.13 &  
$(0.13,\,0.00)$  \\
\hline
$ e^+ \, p \to e^+ \, X$ & ZEUS NC \hfill \cite{Chekanov:2001qu} & 227$\,$ & 1.69 &  
$(0.45,\mathbf{3.13})$ $(0.42,\,0.32)$ $(0.35,\mathbf{3.20})$ $(0.29,\,0.80)$ $(0.18,\,0.64)$  \\
$ e^+ \, p \to e^+ \, X$ & ZEUS NC \hfill \cite{Chekanov:2003yv} & 90$\,$ & 0.36 &  
$(0.22,\,0.01)$ $(0.14,\,1.61)$  \\
$ e^+ \, p \to \nu \, X$ & ZEUS CC \hfill \cite{Breitweg:1999aa} & 29$\,$ & 0.55 &  
$(0.55,\,0.04)$  \\
$ e^+ \, p \to \bar{\nu} \, X$ & ZEUS CC \hfill \cite{Chekanov:2003vw} & 30$\,$ & 0.32 &  
$(0.32,\,0.10)$  \\
$ e^- \, p \to \nu \, X$ & ZEUS CC \hfill \cite{Chekanov:2002zs} & 26$\,$ & 0.12 &  
$(0.12,\,0.02)$  \\
\hline
$ \mu \, p \to \mu \, X$ & BCDMS $F_2$p \hfill \cite{Benvenuti:1989rh} & 339$\,$ & 2.21 &  
$(0.68,\,0.50)$ $(0.63,\,1.63)$ $(0.43,\,0.80)$ $(0.34,\mathbf{4.93})$ $(0.13,\,0.94)$ \\
$ \mu \, d \to \mu \, X$ & BCDMS $F_2$d \hfill \cite{Benvenuti:1989fm} & 251$\,$ & 0.90 &  
$(0.32,\,0.67)$ $(0.24,\,2.49)$ $(0.19,\,2.09)$ $(0.16,\mathbf{5.22})$ \\
$ \mu \, p \to \mu \, X$ & NMC $F_2$p \hfill \cite{Arneodo:1996qe} & 201$\,$ & 0.49 &  
$(0.20,\mathbf{4.56})$ $(0.17,\mathbf{4.76})$ $(0.12,\,0.50)$  \\
$ \mu \, p/d \to \mu \, X$ & NMC $F_2$p/d \hfill \cite{Arneodo:1996qe} & 123$\,$ & 2.17 &  
$(0.61,\,1.11)$ $(0.56,\mathbf{3.60})$ $(0.43,\,0.90)$ $(0.36,\,0.79)$ $(0.21,\,1.41) \;$  \\
\hline
$ p \, \mathrm{Cu} \to \mu^+\mu^-X$ & E605  \hfill \cite{Moreno:1990sf} & 119$\,$ & 1.52 &  
$(0.91,\,1.29)$ $(0.38,\,1.12)$ $(0.23,\,0.31)$  \\
$ pp,pd \to \mu^+ \mu^- \, X$ & E866 pp/pd \hfill \cite{Towell:2001nh} & 15$\,$ & 1.92 &  
$(0.88,\,0.57)$ $(0.69,\,1.15)$ $(0.35,\,1.80)$  \\
$ pp \to \mu^+ \mu^- \, X$ & E866 pp \hfill \cite{Webb:2003ps} & 184$\,$ & 1.52 &  
$(0.75,\,0.04)$ $(0.39,\,1.79)$ $(0.23,\,1.94)$ $(0.14,\mathbf{3.57})$  \\
\hline
$\; \bar{p} p \to (W\!\to \ell \nu) X$ & CDF Wasy \hfill \cite{Abe:1994rj} & 11$\,$ & 0.91 &  
$(0.57,\,0.33)$ $(0.34,\,0.51)$  \\
$ \bar{p} p \to (W\!\to \ell \nu) X$ & CDF Wasy \hfill \cite{Acosta:2005ud} & 11$\,$ & 0.16 &  
$(0.16,\,2.84)$  \\
$ \bar{p} \, p \to \mbox{jet} \, X$ & CDF Jet \hfill \cite{cdfR2} & 72$\,$ & 0.92 &  
$(0.48,\,0.47)$ $(0.44,\mathbf{3.86})$  \\
$ \bar{p} \, p \to \mbox{jet} \, X$ & D0 Jet \hfill \cite{d0R2} & 110$\,$ & 0.68 &  
$(0.39,\,1.70)$ $(0.29,\,0.76)$  \\
\hline
$ \nu \, Fe \to \mu \, X$ &  NuTeV $F_2$ \hfill \cite{Yang:2000ju} & 69$\,$ & 0.84 &  
$(0.37,\,2.75)$ $(0.29,\,0.42)$ $(0.18,\,0.97)$  \\
$ \nu \, Fe \to \mu \, X$ &  NuTeV $F_3$ \hfill \cite{Seligman:1997mc} & 86$\,$ & 0.61 &  
$(0.30,\,0.50)$ $(0.16,\,1.35)$ $(0.15,\,0.30)$  \\
$ \nu \, Fe \to \mu X$ & CDHSW  \hfill \cite{Berge:1989hr} & 96$\,$ & 0.13 &  
$(0.13,\,0.04)$  \\
$ \nu \, Fe \to \mu X$ & CDHSW  \hfill \cite{Berge:1989hr} & 85$\,$ & 0.11 &  
$(0.11,\,1.32)$  \\
$ \nu \, \mathrm{Fe} \to \mu^+ \mu^- \mathrm{X}$ &  NuTeV \hfill \cite{Goncharov:2001qe} & 38$\,$ & 0.68 &  
$(0.39,\,0.31)$ $(0.29,\,0.66)$  \\
$ \bar{\nu} \, \mathrm{Fe} \to \mu^+ \mu^- \mathrm{X}$ & NuTeV \hfill \cite{Goncharov:2001qe} & 33$\,$ & 0.56 &  
$(0.32,\,0.18)$ $(0.24,\,2.56)$  \\
$ \nu \, \mathrm{Fe} \to \mu^+ \mu^- \mathrm{X}$ & CCFR \hfill \cite{Goncharov:2001qe} & 40$\,$ & 0.41 &  
$(0.24,\,1.37)$ $(0.17,\,0.12)$  \\
$ \bar{\nu} \, \mathrm{Fe} \to \mu^+ \mu^- \mathrm{X}$ & CCFR \hfill \cite{Goncharov:2001qe} & 38$\,$ & 0.14 &  
$(0.14,\,0.79)$  \\
\hline
\end{tabular}
\vskip -20pt
\renewcommand\arraystretch{1}
  \end{center}
  \caption{Experiments in the PDF fit that provide at least one measurement with 
$\gamma_i > 0.1\,$.  Large discrepancies ($\sigma_i > 3$) are shown in boldface.
  \label{table:table1}}
\end{table*}
\endgroup

\emph{The centerpiece of this study is presented in Table \ref{table:table1},} 
which lists all of the measurements $(\gamma_i,\sigma_i)$ that pass the 
importance criterion $\gamma_i \, > \, 0.1\,$.  The parameter 
$\gamma_i$ measures the importance of the experiment under study in 
determining the result $z_i = 0$ of the global fit, while $\sigma_i$ measures 
the discrepancy between that experiment and the consensus of the others.  
One must keep in mind that to generate this table, the DSD method had to be 
applied separately for each experiment.  Hence the definition of the 
coordinates $z_i$ is different for each line in the table. The measurements 
from each data set are listed in descending order of $\gamma_i$, so in each 
case, $i = 1$ labels the parameter that is measured best by the experiment 
under study.  

The data sets in Table \ref{table:table1} are grouped according to their 
initial-state particles.
These groupings involve different experimental techniques, and even different 
laboratories.  The $e p$ results from HERA (H1+ZEUS) cover similar kinematic 
regions using similar techniques, but they are listed separately to satisfy 
possible curiosity.  From a theoretical point of view, $e p \to e X$ and 
$\mu p \to \mu X$ deep inelastic scattering (DIS) measurements are 
equivalent.  However, the $\mu p$ data are from fixed-target experiments that 
cover a different kinematic region from the $e p$ experiments, as will be 
discussed in Sec.\ \ref{sec:MuonExpts}.  

\begin{table}[htb]
\vskip 10pt
  \begin{center}
\null
\begin{tabular}{||c|c|r||}
\hline
Process & Expt & N \\
\hline
$ e^- \, p \to e^- \, X$ & H1 NC \hfill \cite{Adloff:2003uh} & 13   \\ 
$ e^+ \, p \to \nu \, X$ & H1 CC \hfill \cite{Adloff:2003uh} & 28  \\ 
$ e^+ \, p \to c \,X$ & H1 $F_2^{\,c}$ \hfill \cite{Adloff:2001zj} & 8  \\ 
$ e^+ \, p \to c \, X$ & H1 $F_2^{\,c}$ \hfill \cite{Aktas:2005iw,Aktas:2004az} & 10  \\ 
$ e^+ \, p \to b \, X$ & H1 $F_2^{\,b}$ \hfill \cite{Aktas:2005iw,Aktas:2004az} & 10  \\ 
\hline
$ e^- \, p \to e^- \, X$ & ZEUS NC \hfill \cite{Chekanov:2002ej} & 92   \\ 
$ e^+   \, p \to c \bar{c} X$ & ZEUS $F_2^{\,c}$ \hfill \cite{Breitweg:1999ad} & 18   \\ 
$ e^\pm \, p \to c \bar{c} X$ & ZEUS $F_2^{\,c}$ \hfill \cite{Chekanov:2003rb} & 27  \\  
\hline
\end{tabular}
\vskip -10pt
  \end{center}
  \caption{Experiments in the PDF fit with no measurements with 
$\gamma_i > 0.1$.
  \label{table:table2}}
\end{table}

In addition to the 29 data sets listed in Table \ref{table:table1}, the fit 
includes the 8 data sets listed in Table \ref{table:table2} which contribute 
no information of importance $\gamma_i > 0.1\,$.  
These are all HERA experiments with relatively low statistics.

Table \ref{table:table1} shows that the HERA data contribute a 
substantial portion of our knowledge on PDFs.  However, it also shows 
major contributions from fixed-target $\mu p$ and $\mu d$ DIS 
experiments---not surprisingly, because of their high statistics, and because 
the deuterium target measurements help to differentiate among quark flavors.  
There are also major contributions from Drell-Yan (DY) lepton pair production 
on fixed targets; from Tevatron $\bar{p} p$ inclusive jet experiments and 
the forward-backward lepton asymmetry from W decay; and
from neutrino experiments.  

It is shown in \cite{DSD} that $\gamma_i$ can be interpreted as the fraction of 
the global measurement $z_i = 0 \pm 1$ that is contributed by the data in 
$\mathbf{S}$.  The column listing $\sum_i \gamma_i$ in Table \ref{table:table1} 
can therefore be thought of as the number of fitting parameters that are 
determined by the experiment in question.  Totaling these for each experimental 
category, we find that 
H1 and ZEUS experiments combined effectively measure 6.2 parameters;
$\mu p$ experiments measure 5.8; DY experiments measure 5.0; 
neutrino experiments measure 3.5; and 
Tevatron experiments measure 2.7.  
The sum of these numbers is 23.2, which is satisfyingly close to the actual 
number $n = 24$ of parameters that were fitted.
\emph{The fact that all of these 
types of experiment are needed to get the best information on 
PDFs has long been believed; but it is established here quantitatively for the 
first time.}

\begin{figure}[tbh]
\vskip 10pt
\begin{center}
 \resizebox*{0.45\textwidth}{!}{
\includegraphics[clip=true,scale=1.0]{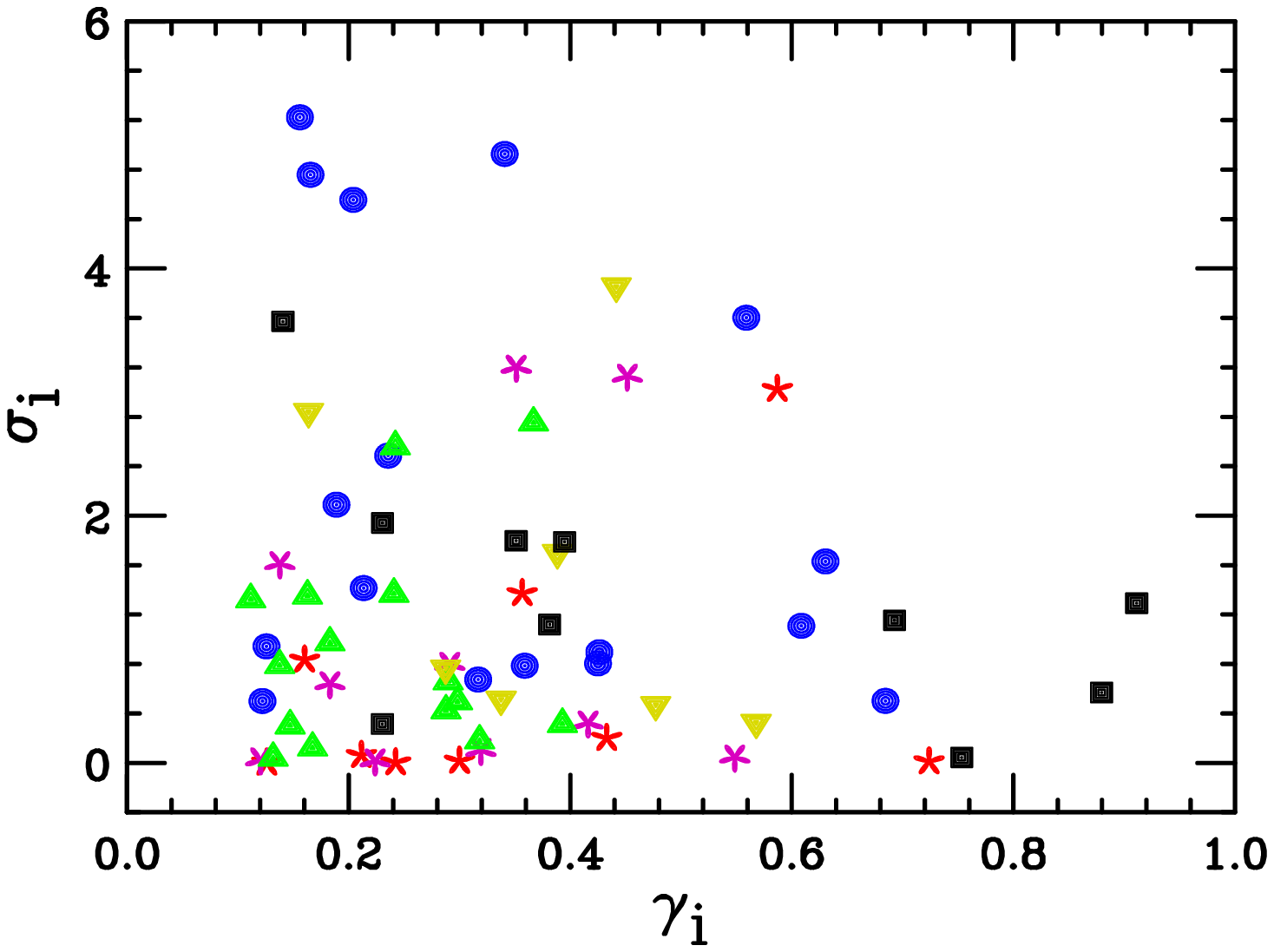}}
\end{center}
\vskip -15pt
 \caption{Results from Table \ref{table:table1}:
$e p$ (daisy); 
$\mu p, \mu d$ ({\Large $\circ$});
$p p, p d, p\mathrm{Cu}$ ($\Box$);
$\bar{p} p$ ($\nabla$);
$\nu A$ ($\Delta$).
}
 \label{fig:figOne}
\vskip 5pt
\end{figure}

The results in Table \ref{table:table1} are displayed graphically in 
Fig.\ \ref{fig:figOne}.  This plot shows that the effective measurements 
are widely distributed in the $(\gamma,\sigma)$ plane. 
Broadly speaking, all of the experiment types contribute to all parts 
of the plot, with one possible exception that is explored in 
Sec.\ \ref{sec:MuonExpts}.
Smaller values of $\gamma$ are more common because most aspects of 
the fit are constrained by more than one experiment.
Smaller values of $\sigma$ are more common because the fit is 
reasonably self-consistent.  The distribution of $\sigma$ is examined in 
detail in the next Section.

\section{Distribution of the discrepancies
\label{sec:Distribution}}

\begin{figure}[tbh]
\vskip 10pt
\begin{center}
 \resizebox*{0.40\textwidth}{!}{
\includegraphics[clip=true,scale=1.0]{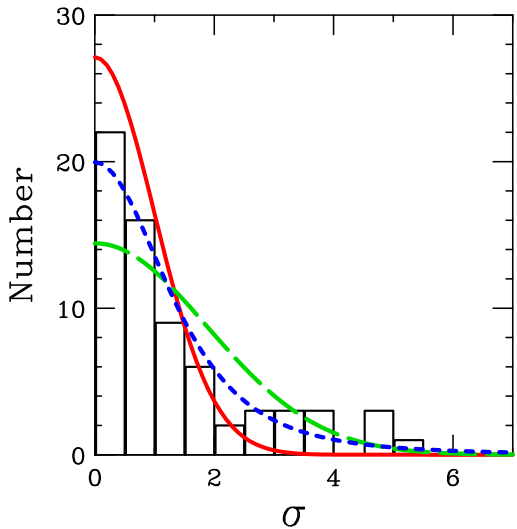}}
\end{center}
\vskip -15pt
 \caption{Distribution of the discrepancies $\sigma_i$ from 
Table \ref{table:table1}. The solid curve is the parameter-free Gaussian 
prediction (\ref{eq:AbsoluteGaussian}).  The long-dashed curve is a 
fit to the scaled Gaussian form (\ref{eq:ScaledGaussian}).  The short-dashed 
curve is a fit to the squared-Lorentzian form (\ref{eq:SquaredLorentzian}).
}
 \label{fig:figTwo}
\vskip 5pt
\end{figure}

According to Gaussian statistics, the 68 discrepancies $\{\sigma_i\}$ 
listed in Table \ref{table:table1} would be expected to follow the normal 
distribution
\begin{equation}
\frac{dP}{d\sigma} \, = \, 
\sqrt{\frac{1}{2 \pi}} \, \exp(- \sigma^2/2) \; .
\label{eq:AbsoluteGaussian}
\end{equation}
A histogram of the actual distribution is shown in Fig.\ \ref{fig:figTwo},
together with that prediction.
The distribution is clearly broader than the prediction.
\emph{Hence, the observed inconsistencies among the data 
sets are larger than what is predicted by Gaussian statistics.}  
This can also be seen from the number of ``outliers:''
10 measurements out of 68 in Table \ref{table:table1} 
have $\sigma_i > 3\,$. 
The probability for so many large 
values to arise by random fluctuations from the distribution
(\ref{eq:AbsoluteGaussian}) is vanishingly small---even 5
instances of $|\sigma_i| > 3$ in 68 tries is a million-to-one long shot.

When it is necessary to combine experimental results that lie 
outside a comfortable range of statistical agreement, a standard
course of action is to scale up the errors---see, \textit{e.g.}, the Particle 
Data Group tables in \cite{PDG}.
That approach suggests fitting the histogram in Fig.\ \ref{fig:figTwo} 
to a Gaussian form with adjustable width:
\begin{equation}
\frac{dP}{d\sigma} \, = \, 
\sqrt{\frac{1}{2 \, \pi \, c^2 \,}} \, 
\exp(- \sigma^2/(2\,c^2)) \; .
\label{eq:ScaledGaussian}
\end{equation}
A maximum-likelihood fit to this form yields $c = 1.88\,$.  
\emph{This suggests that the errors in the PDF fit need to be scaled up 
by nearly a factor of 2 to allow for the observed inconsistencies
among the data sets.}  
This fit is also shown in Fig.\ \ref{fig:figTwo}. 

Although the scaled Gaussian is an improvement over the absolute one, 
the fit it provides is not entirely satisfactory. 
A much better description of the histogram can be obtained using a form 
with a more slowly falling tail, such as the squared-Lorentzian:
\begin{equation}
\frac{dP}{d\sigma} \, = \, 
\frac{2\,m^3/\pi}{(\sigma^2 \, + \, m^2)^2} \; .
\label{eq:SquaredLorentzian}
\end{equation}
This curve is also shown in Fig.\ \ref{fig:figTwo}, using the 
parameter value $m = 2.17$ obtained by maximum-likelihood fitting.

\begin{figure}[tbh]
\vskip 10pt
\begin{center}
\hfill
 \resizebox*{0.20\textwidth}{!}{
\includegraphics[clip=true,scale=1.0]{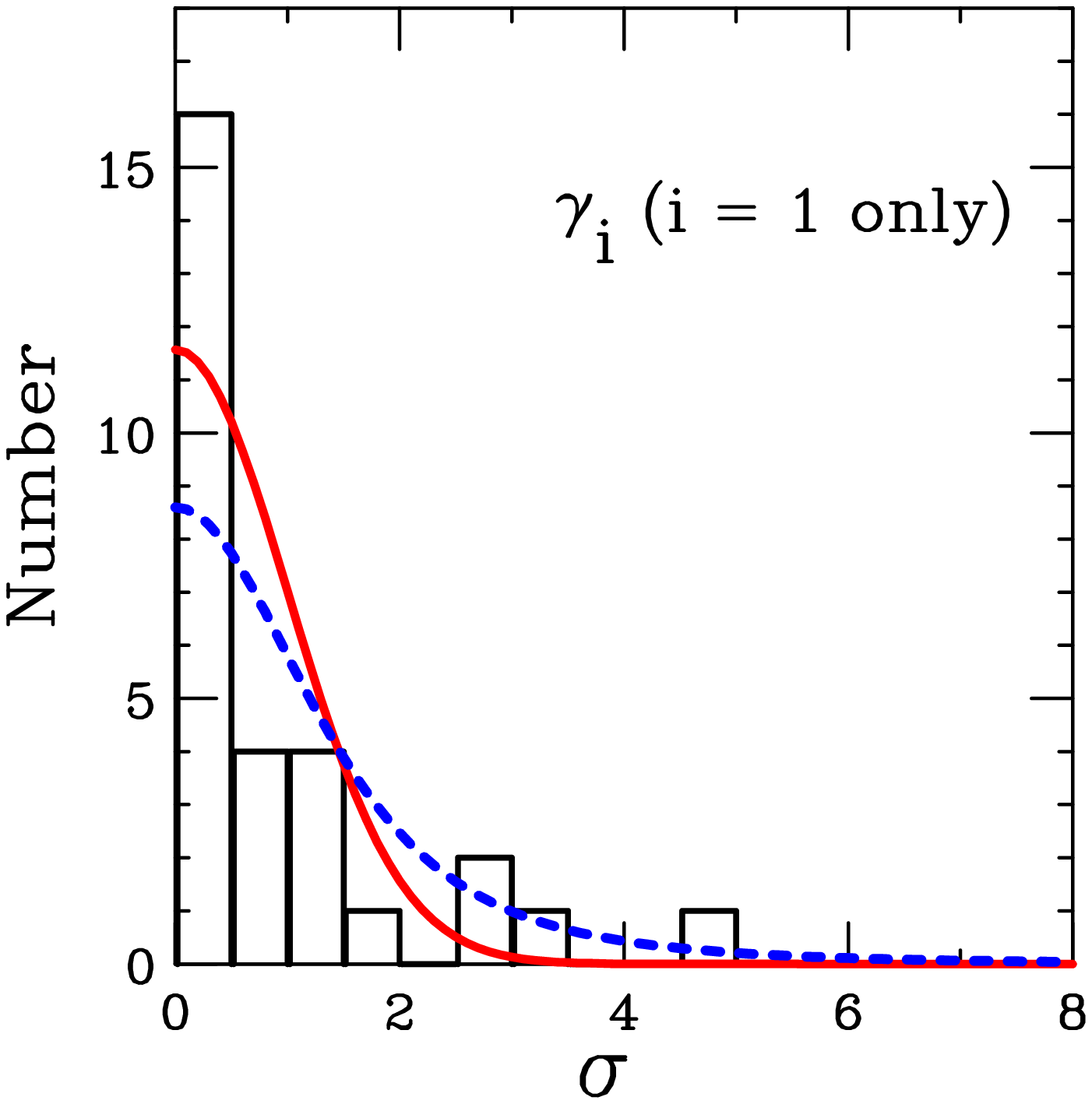}}
\hfill
 \resizebox*{0.20\textwidth}{!}{
\includegraphics[clip=true,scale=1.0]{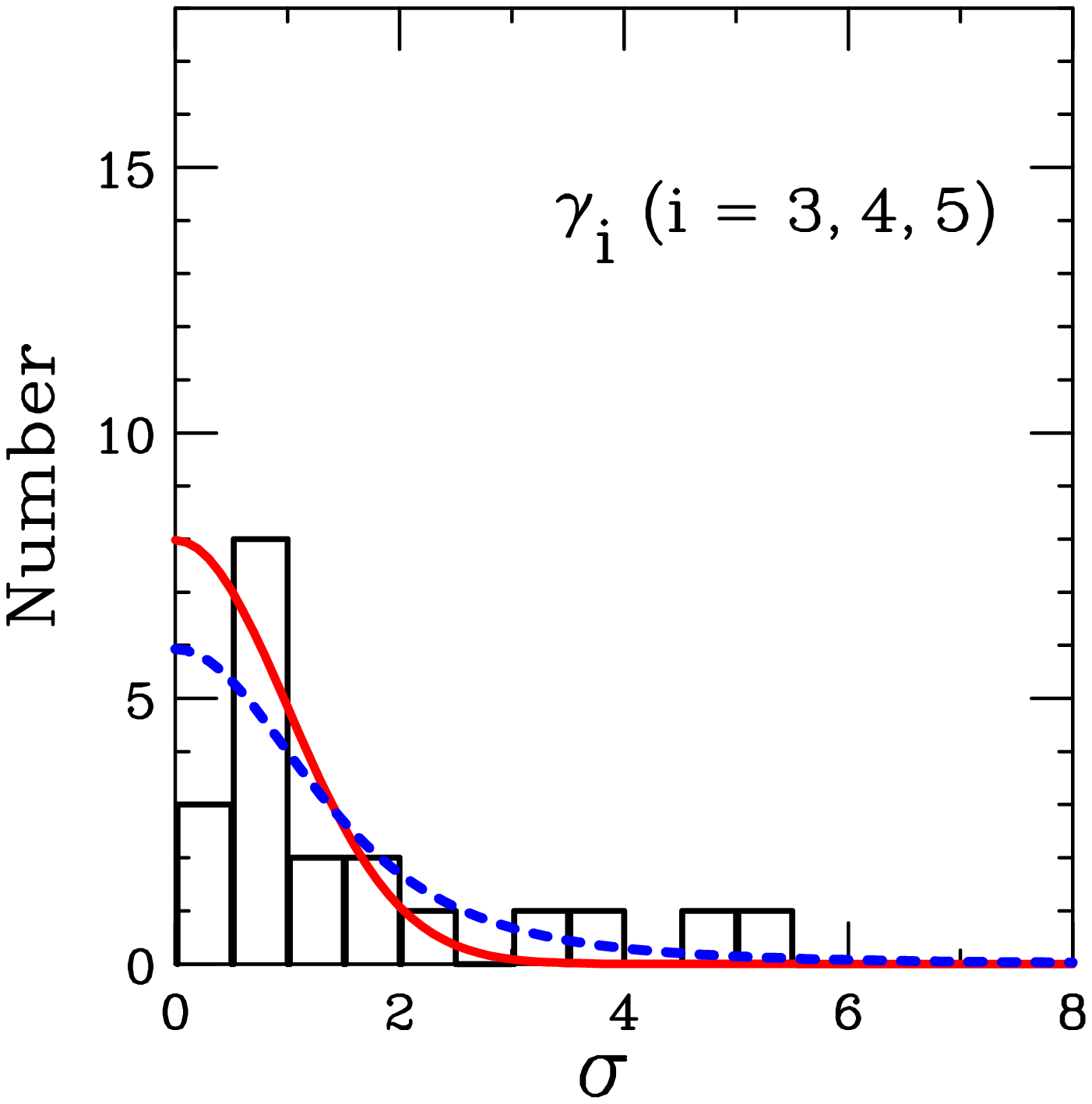}}
\hfill
\end{center}
\begin{center}
\hfill
 \resizebox*{0.20\textwidth}{!}{
\includegraphics[clip=true,scale=1.0]{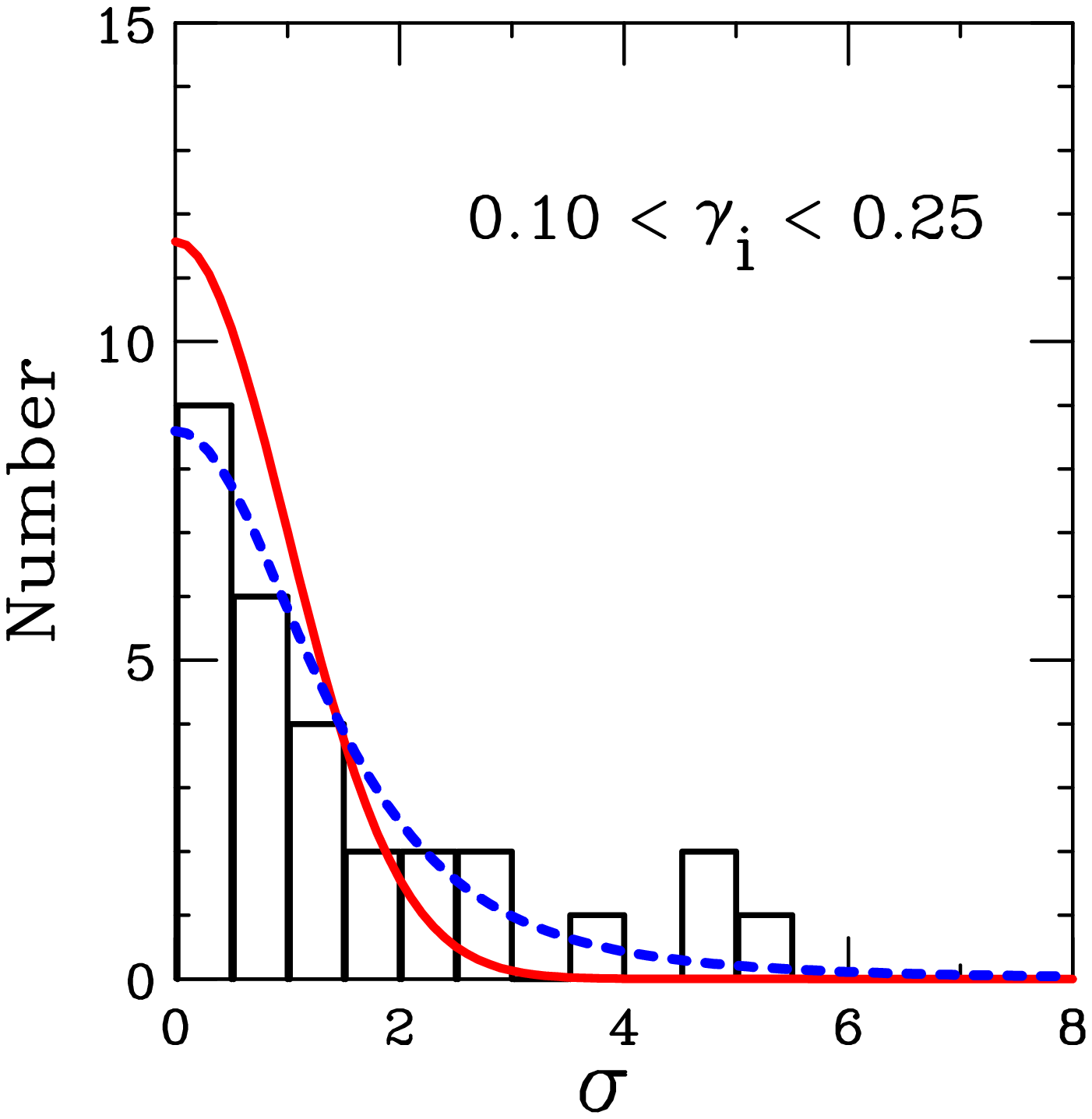}}
\hfill
 \resizebox*{0.20\textwidth}{!}{
\includegraphics[clip=true,scale=1.0]{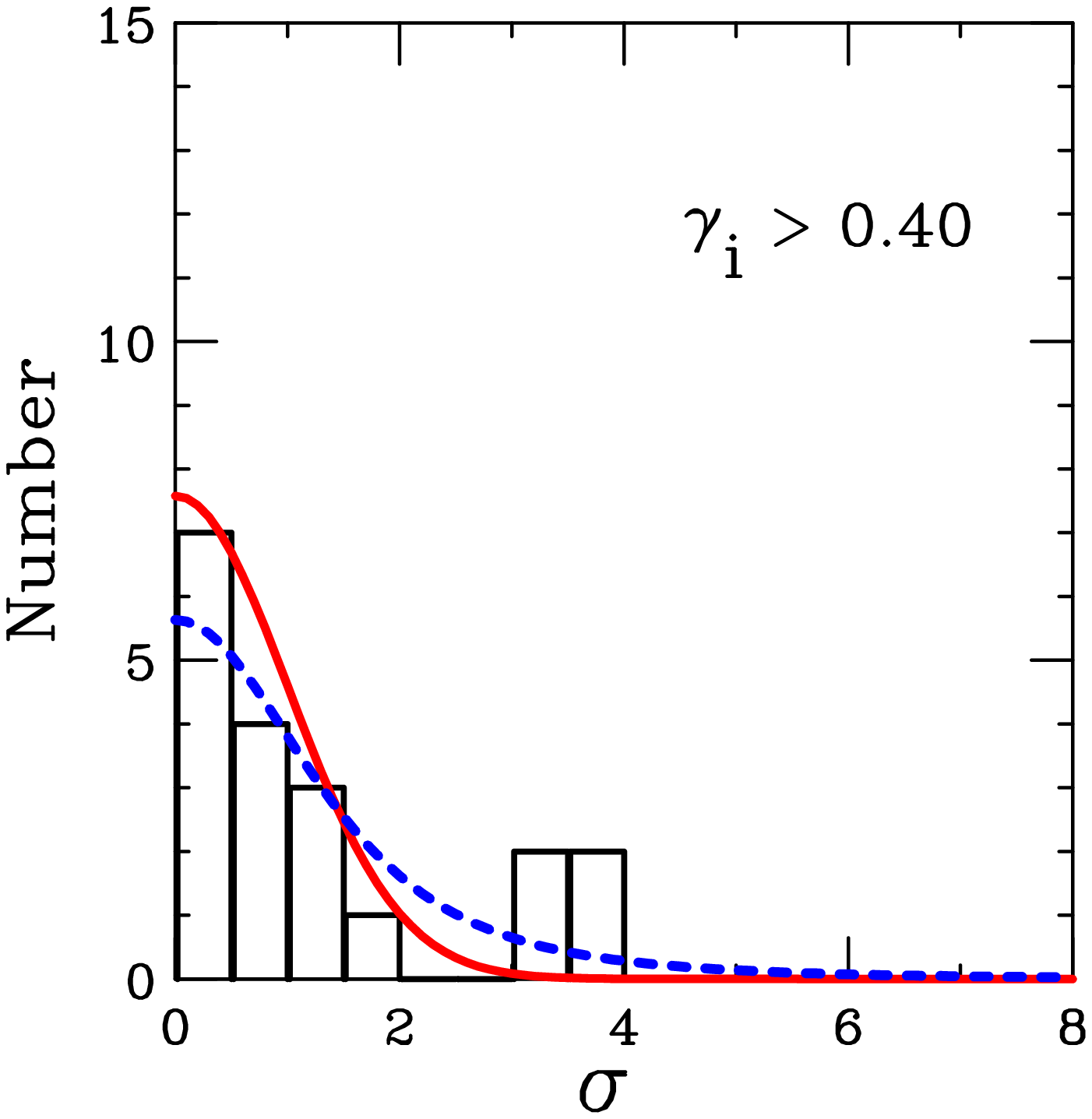}}
\hfill
\end{center}
\vskip -15pt
 \caption{Distribution of the discrepancies $\sigma_i$ from 
Table \ref{table:table1} for measurements with:
(a) $i=1$ only; (b) $i=3,4,5$; (c) $0.10 < \gamma_i < 0.25$;
(d) $\gamma_i > 0.40\,$.  Curves shown are the absolute 
Gaussian and squared-Lorentzian curves from 
Fig.\ \ref{fig:figTwo}, normalized to the number of 
points in each histograms, but \emph{not} refitted.
}
 \label{fig:figThree}
\vskip  5pt
\end{figure}

The distribution of discrepancies seen in Fig.\ \ref{fig:figTwo} appears to 
be a general characteristic of the global fit.  This is demonstrated by 
Fig.\ \ref{fig:figThree}, which shows histograms for various 
subsets of the $(\gamma_i,\sigma_i)$ pairs from 
Table \ref{table:table1}: 
(a) those with $i=1$, \textit{i.e.}, the best-measured parameter from each 
experiment; 
(b) those with $i=3,4,5$, \textit{i.e.}, less well-measured parameters from 
each experiment;
(c) those with $0.10 < \gamma_i < 0.25$, \textit{i.e.}, parameters that are 
weakly determined by the experiment under study; and 
(d) those with $\gamma_i > 0.40$, \textit{i.e.}, parameters that are strongly 
determined by the experiment under study.
(The middle ranges---$\, i\!=\!2$ in (a) and (b), 
$0.25 \!< \!\gamma_i \!< \! 0.40$ in (c) and (d)---are excluded from these 
histograms in an attempt to accentuate any systematic differences.)
As far as can be seen with the limited statistics, these 
distributions all look alike.  They are all \emph{inconsistent} with the 
absolute Gaussian prediction, and they are all \emph{consistent} with 
the squared-Lorentzian form, whose width parameter $m=2.17$ is 
kept the same as in Fig.\ \ref{fig:figTwo}.

The only systematic trend that is suggested by Fig.\ \ref{fig:figOne} is 
a tendency for the muon experiments to have larger-than-average 
discrepancies.  That trend is explored in the next Section.

\section{Role of the muon experiments}
\label{sec:MuonExpts}

Figure \ref{fig:figOne} (or Table \ref{table:table1}) shows that the four 
largest discrepancies $\sigma_i$ all come from the $\mu p$ and $\mu d$ 
fixed-target BCDMS and NMC experiments.  This is perhaps not surprising, 
since a significant tension between those experiments and the rest of the 
global fit was already observed in CTEQ5 \cite{CTEQ5}, using the less 
sophisticated method of plotting $\chi_{\mathbf{S}}^{\, 2}$ vs.\ 
$\! \chi_{\mathbf{\overline{S}}}^{\, 2}$ \cite{Collins}.  Tension between 
the NMC and BCDMS data sets can also be inferred from a recent MSTW 
paper \cite{MSTWalphas}, which shows that the two experiments prefer values 
of $\alpha_s(m_Z)$ that differ significantly, in opposite directions, from 
the approximate world-average value $0.118$ that is used here \cite{PDG}.

\begin{figure*}[tbh]
\vskip 10pt
\begin{center}
 \resizebox*{0.40\textwidth}{!}{
\includegraphics[clip=true,scale=1.0]{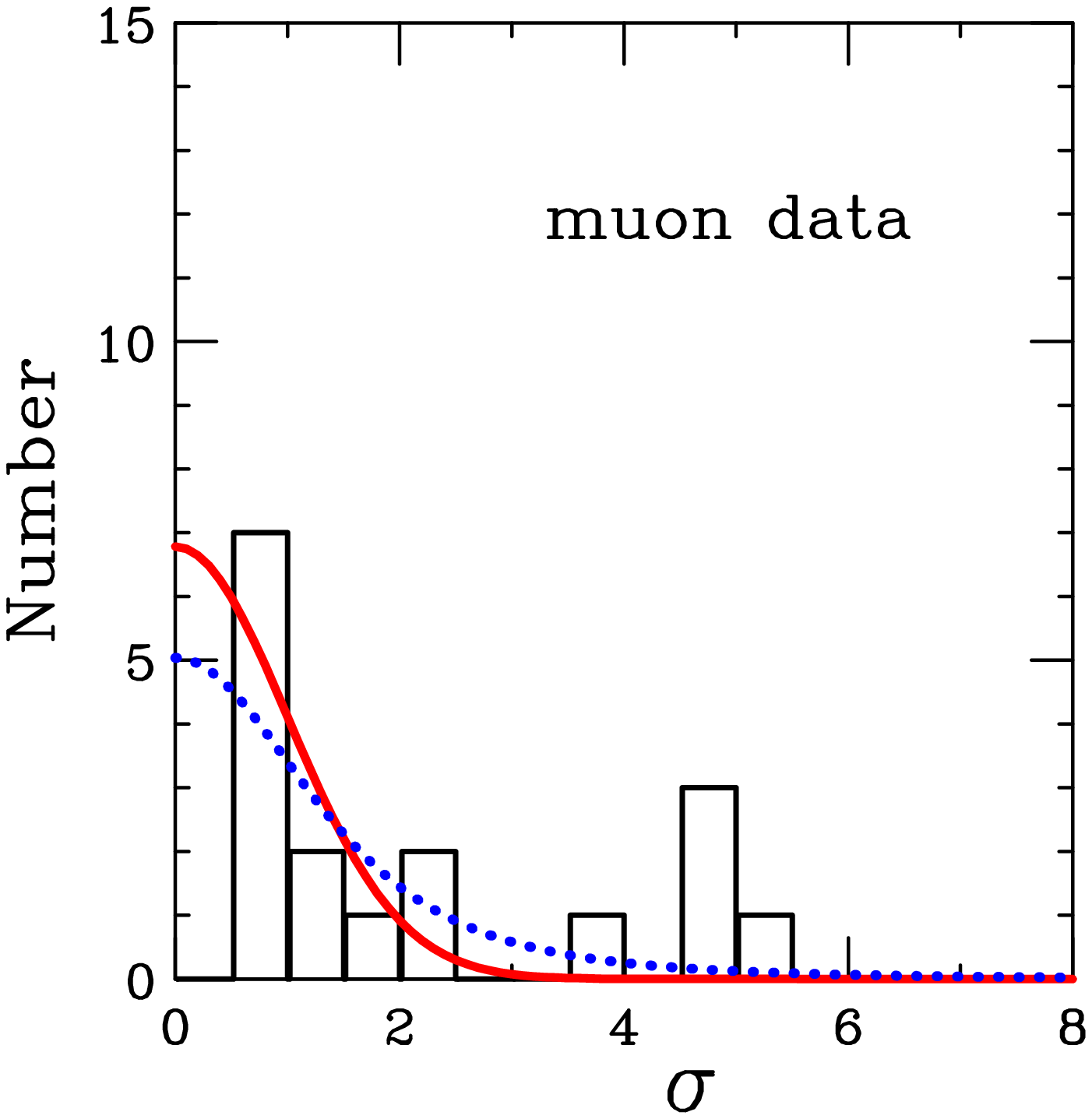}}
 \resizebox*{0.40\textwidth}{!}{
\includegraphics[clip=true,scale=1.0]{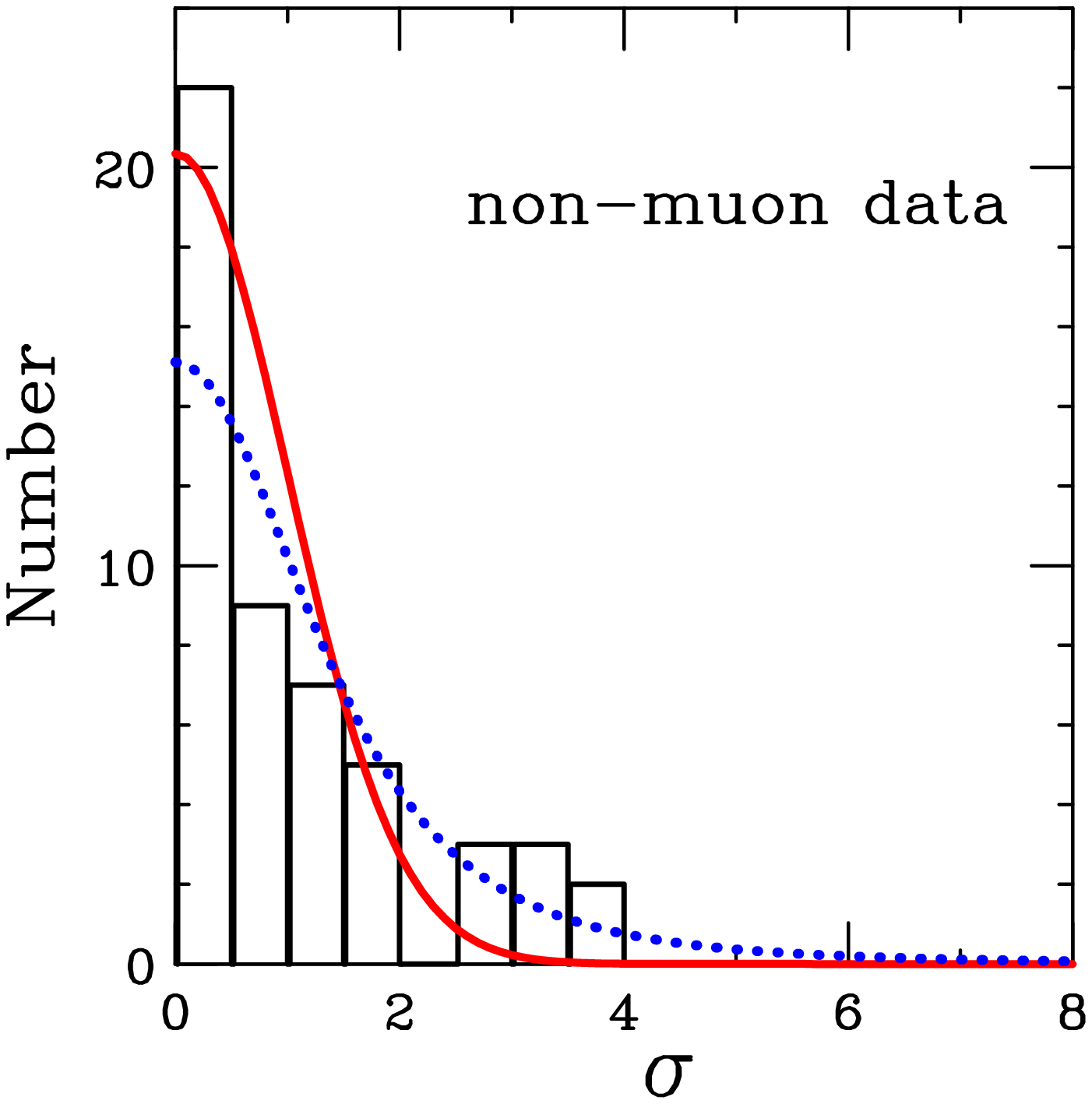}}
\end{center}
\vskip -15pt
 \caption{Distribution of the discrepancies $\sigma_i$ from 
Table \ref{table:table1}.  Left panel: $\mu p$ and $\mu d$ experiments only.  
Right panel: all experiments except $\mu p$ and $\mu d$.  Curves are the 
absolute Gaussian prediction and the squared-Lorentzian fit---both the same 
as shown in Fig.\ \ref{fig:figThree} except for normalization.
}
 \label{fig:figFour}
\vskip  5pt
\end{figure*}

Figure \ref{fig:figFour} shows the histogram of $\sigma_i$ for the muon 
experiments and the others separately.  The muon histogram looks quite 
different: it contains zero counts in the first bin and all four of the 
counts with $\sigma > 4$.  It is therefore natural to raise the question 
of whether some or all of the muon experiments---or their theoretical 
treatment---may contain important systematic errors that have been
neglected.  This question is explored in detail in Sec.\ \ref{sec:MoreMuon}.

\begin{figure}[tbh]
\vskip 10pt
\begin{center}
 \resizebox*{0.40\textwidth}{!}{
\includegraphics[clip=true,scale=1.0]{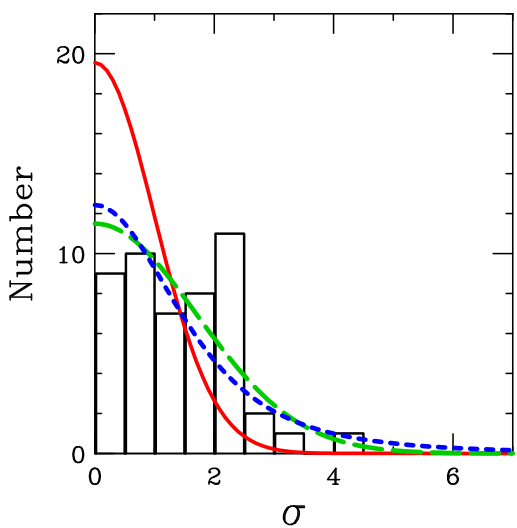}}
\end{center}
\vskip -15pt
 \caption{Distribution of the discrepancies $\sigma_i$ from global fits 
with the $\mu p$ data omitted.  Solid curve is the parameter-free Gaussian 
prediction (\ref{eq:AbsoluteGaussian}).  Long dash and short dash curves are 
new fits to scaled-Gaussian (\ref{eq:ScaledGaussian}) and squared-Lorentzian 
(\ref{eq:SquaredLorentzian}) forms respectively.
}
 \label{fig:figFive}
\vskip  5pt
\end{figure}

The essential question from the standpoint of this paper is whether or not
the deviation from Gaussian behavior seen in Fig.~\ref{fig:figTwo} is a 
general characteristic of the PDF fits, or whether it could instead just 
point to problems with the muon data sets.  The right hand side of 
Fig.\ \ref{fig:figFour} appears to show that the non-muon data also have 
a tail at large $\sigma$ which is inconsistent with the ideal Gaussian 
curve.  However, if one speculates that the muon data or their theoretical 
treatment may be incorrect, then those large-$\sigma$ points might merely 
reflect a conflict with the muon data.  A direct way to proceed is to 
repeat the analysis that led to Table \ref{table:table1}, with all 
four of the muon experiments omitted from the global fit.  
In carrying this out, it was necessary 
to reduce the number of fitting parameters from 24 to 21 in order to 
obtain stable fits to the reduced input data. Results from this 
study are shown in Fig.\ \ref{fig:figFive}.  \emph{The distribution is 
again broader than the absolute Gaussian prediction,} so 
the central conclusion from Sec.\ \ref{sec:Distribution} stands
even if the muon data are excluded. 
Because no extreme outlying points appear in this histogram, a rescaled 
Gaussian ($c \! = \! 1.70$ in Eq.~(\ref{eq:ScaledGaussian})) this time 
gives an acceptable fit.  It is even slightly better than a 
squared-Lorentzian fit ($m = 2.51$ in Eq.~(\ref{eq:SquaredLorentzian})).

The DSD method can be used to further explore the contribution of 
specific experiments to the global fit.  This is pursued for the muon 
experiments in Sec.\ \ref{sec:MoreMuon}.

\section{Non-Gaussian statistics and $\mathbf{\Delta\chi^2}$
\label{sec:NonGaussian}}
This Section explains the concept of $\Delta\chi^2$ and estimates 
it for the PDF fit.  Let us consider a simple scenario that is similar to 
the DSD situation of measuring a single variable $z_i$ in the symmetric case 
$\gamma_i = 0.5\,$.  Specifically, suppose a quantity $z$ is measured by two 
equally-trustworthy experiments, which report
\begin{eqnarray}
  \mbox{Expt 1:}\; \; z &=& A \, \pm \, \sqrt{2} \nonumber  \\
  \mbox{Expt 2:}\; \; z &=& B \, \pm \, \sqrt{2} \; .
  \label{eq:AB} 
\end{eqnarray}
We wish to combine these two measurements into a single result.  According to 
standard Gaussian statistics, that is done by taking the average and combining 
the errors in quadrature:
\begin{equation}
    z = (A+B)/2 \, \pm \, 1  \; .
  \label{eq:ans} 
\end{equation}
The $\chi^2$ measure of fit quality is
\begin{eqnarray}
  \chi^2 &=& \left(\frac{z-A}{\sqrt{2}}\right)^2 \, + \, 
             \left(\frac{z-B}{\sqrt{2}}\right)^2 \label{eq:chisq1} \\
         &=& \left(z-\frac{A+B}{2}\right)^2 \, + \, 
             \left(\frac{A-B}{2}\right)^2 \, .  \label{eq:chisq2} 
\end{eqnarray}
The algebraic rearrangement in Eq.\ (\ref{eq:chisq2}) reveals that the expected 
best-fit value $z = (A+B)/2$ indeed minimizes $\chi^2$, and 
that the error limits in (\ref{eq:ans}) correspond to the points where 
$\chi^2 = \chi_{\mathrm{min}}^{\, 2} + \Delta\chi^2$ with $\Delta\chi^2 = 1$.  
This corresponds to the 68.3\% confidence limit, \textit{i.e.}, 
``$1 \, \sigma$.''  The uncertainty limit for 90\% confidence is farther from 
the minimum in $z$ by a factor $1.64$, which corresponds to 
$\Delta\chi^2 = 2.71\,$. 

Now let us see what happens if we do not assume that the errors are 
Gaussian.  Suppose instead that the measurements $A$ and $B$ 
arise from independent random processes with probability distributions
\begin{eqnarray}
    \frac{dP}{dA} &=& f(A) \\
    \frac{dP}{dB} &=& f(B) \; , 
  \label{eq:dPdAdPdB} 
\end{eqnarray}
where $\int_{-\infty}^{\infty} f(A) \, dA = 1$, and we assume for simplicity 
that $A$ and $B$ come from the same distribution.  We can assume without loss 
of generality that this distribution is centered about a true answer of $0$. 
Let us also assume that the distribution is symmetric: $f(A) = f(-A)$.  
It is intuitively clear that the best estimate from the two measurements will
remain equal to the average $(A+B)/2$, so the real issue is how to assess the 
uncertainty on that result.

If the two measurements can be repeated many times, the probability 
distribution for their average $(A+B)/2$ is given by
\begin{eqnarray}
    \frac{dP}{dz} &=& \rho_z(z) \nonumber \\
    &=&  
   \int_{-\infty}^\infty \! dA \, f(A) 
   \int_{-\infty}^\infty \! dB \, f(B) \;
   \delta \! \left(\frac{A+B}{2} - z\right) \nonumber \\
   &=& 
   2 \int_{-\infty}^\infty \! dA \, f(A) \, f(2z - A) \; .
  \label{eq:dPdz} 
\end{eqnarray}
Meanwhile, the probability distribution for the difference
$(A - B)$ between the two measurements, expressed in units of 
its error $\sqrt{(\sqrt{2})^2 + (\sqrt{2})^2} = 2$, is given by
\begin{eqnarray}
    \frac{dP}{d\sigma} &=& \rho_\sigma(\sigma) \nonumber \\
   &=& 
   \int_{-\infty}^\infty \! dA \, f(A) 
   \int_{-\infty}^\infty \! dB \, f(B) \;
   \delta \! \left(\frac{A-B}{2} - \sigma\right) \nonumber \\
   &=&
   2 \int_{-\infty}^\infty \! dA \, f(A) \, f(A - 2\sigma) \; .
  \label{eq:dPdsigma} 
\end{eqnarray}
Comparing Eqs.~(\ref{eq:dPdz}--\ref{eq:dPdsigma}) and using 
the assumed symmetry $f(A) = f(-A)$, we obtain 
\begin{equation}
    \rho_z(z) \, = \, \rho_\sigma(z)  \; .
  \label{eq:rhoz} 
\end{equation}
\emph{Eq.~(\ref{eq:rhoz}) shows that the uncertainty distribution for the 
average of the two measurements is the same as the uncertainty distribution 
for their difference}, when that difference is normalized by its error as 
is done here.  The former is what is needed to estimate the uncertainties of 
the PDF results, while the latter is what is measured in the histogram of 
Fig.\ \ref{fig:figTwo}.  

Before proceeding, let us check that the above formulae reproduce the
correct results in the Gaussian case $f(A) = \exp(-A^2/4)/\sqrt{4\,\pi}$. 
In that case, Eqs.~(\ref{eq:dPdz}) and (\ref{eq:rhoz}) give 
$\rho_z(z) = \exp(-z^2/2)/\sqrt{2 \, \pi}$ and 
$\rho_\sigma(\sigma) = \exp(-\sigma^2/2)/\sqrt{2\, \pi}$.  
Thus $dP/dz$ and $dP/d\sigma$ are both Gaussians of width 1,
which indeed agrees with the standard rules for propagating 
the uncertainties from Eq.\ (\ref{eq:AB}).
The middle 68.3\% (90\%) of the probability distribution 
$dP/dz$ is contained in $|z| < 1.00$ ($1.64$), which 
corresponds to the points where $\Delta\chi^2 = 1.00$ 
($2.71$), in agreement with earlier statements.

If the distribution of differences, and hence according to
(\ref{eq:rhoz}) the distribution of averages, is given by
the scaled Gaussian form (\ref{eq:ScaledGaussian}), then the
uncertainty limits in $z$ are scaled by the parameter $c$
in that formula.  Hence the 90\% confidence tolerance 
becomes $\Delta\chi^2 = 2.71\,c^2$.  For the value
$c = 1.88$ found in Sec.\ \ref{sec:Distribution} from the
fit in Fig.~\ref{fig:figTwo}, this implies $\Delta\chi^2=14\,$. 

If, on the other hand, the distribution of differences, and hence the 
distribution of averages, is given by the squared-Lorentzian form 
(\ref{eq:SquaredLorentzian}) with the width parameter $m=2.17$ that was 
found in Sec.\ \ref{sec:Distribution} by fitting the distribution of 
differences, 
then the central 68.3\% (90\%) of the distribution is contained in 
$|z| < 1.50$ ($2.95$), which 
corresponds to $\Delta\chi^2 = 2.25$ ($\Delta\chi^2 = 8.70$).
Note that the ratio between 68.3\% and 90\% confidence points is 
larger for the squared-Lorentzian distribution ($8.70/2.25 = 3.9$) 
than for the Gaussian distribution ($2.71/1.00 = 2.7$), because of the 
relatively slowly-falling tail of the Lorentzian.

\emph{These results suggest that the 90\% confidence criterion for 
the uncertainty of the global fit is given by $\Delta\chi^2 \approx 10$}.

\section{Remark on $\mathbf{\chi^2/N}$
\label{sec:Remark}}

The overall $\chi_{\mathrm{total}}^{\,2}$ ($ = 3074$) for the global fit is 
not far from the total number of data points ($N_{\mathrm{total}}=2970$) in 
the fit.  This at first seems to contradict the idea that there are 
inconsistencies in the fit that are nearly twice the expectation based on 
the experimental errors.  For example, in the extreme, if the actual errors 
for all of the data points were a factor of 2 larger than the errors claimed 
by the experiments, we would expect 
$\chi_{\mathrm{total}}^2/N_{\mathrm{total}} \approx 4$.

However, the actual situation does not correspond to that extreme.  A given 
experiment with $N$ data points delivers significant information along only a 
few directions in parameter space---at most 5 or 6 according to 
Table \ref{table:table1}.  Of those directions, there is significant discord 
along at most 2 or 3.  We can estimate the effect of this on $\chi^2/N$ as 
follows.  Eq.\ (\ref{eq:BetaGamma}) shows that the lowest possible $\chi^2$ 
for experiment $\mathbf{S}$ occurs at $\{z_i = A_i\}$, while the 
global best-fit value occurs at $\{z_i = 0\}$.  Hence in the global fit, 
$\chi_{\mathbf{S}}^{\, 2}$ lies above its best-fit value by 
\begin{equation}
   \sum_{i=1}^n \left(\frac{A_i}{B_i}\right)^2 \nonumber  \, .
\label{eq:eq1}
\end{equation}
Combining Eqs.\ (\ref{eq:CentralResult1}--\ref{eq:CentralResult2}) 
and using $\, \gamma_i A_i \, + \, (1-\gamma_i) C_i \, = \, 0$ which follows 
from Eq.\ (\ref{eq:diagchi}), we obtain
\begin{eqnarray}
 A_i \, &=& \sqrt{(1-\gamma_i)/\gamma_i} \; \sigma_i \nonumber \\
 B_i \, &=& \sqrt{1/\gamma_i} \; .
\end{eqnarray}
Hence the addition to $\chi^2$ from experiment $\mathbf{S}$ is
\begin{equation}
\sum_{i=1}^n  \left(\frac{A_i}{B_i}\right)^2 \, = \, 
\sum_{i=1}^n \, (1-\gamma_i) \, \sigma_i^{\, 2}. \nonumber
\end{equation}
Adding this up over the 68 $(\gamma_i,\,\sigma_i)$ pairs with $\gamma_i > 0.1$ 
in Table \ref{table:table1} gives a total of $168\,$.  The full data set has 
2970 points, so this contribution of $\approx 168$ to $\chi_{\mathrm{total}}^2$ 
is not large enough to spoil the expectation that 
$\chi_{\mathrm{total}}^2 \approx N_{\mathrm{total}} \pm \sqrt{2\,N_{\mathrm{total}}}$.  

%



\begin{figure}[tbh]
\vskip 10pt
\begin{center}
 \resizebox*{0.45\textwidth}{!}{
\includegraphics[clip=true,scale=1.0]{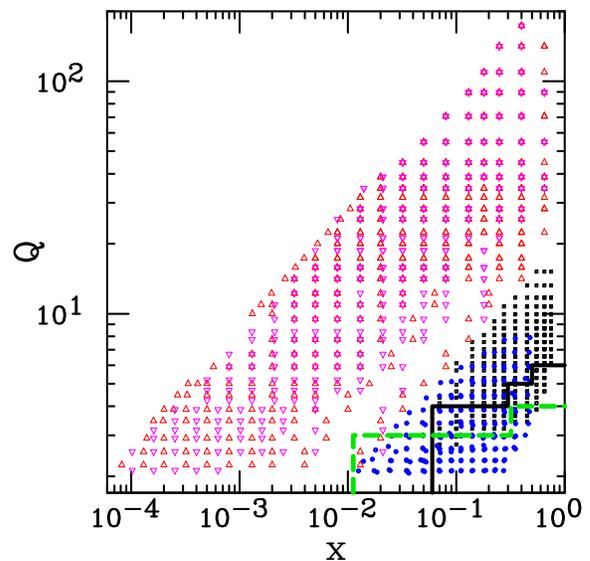}}
\end{center}
\vskip -15pt
 \caption{Kinematic region covered by the lepton DIS experiments
$e p \to e X$ (H1 = $\Delta$, ZEUS = $\nabla$) and
$\mu p \to \mu X$ (BCDMS=$\Box$, NMC = $\circ$).}
 \label{fig:figSix}
\vskip  5pt
\end{figure}

\section{Further study of the muon experiments 
\label{sec:MoreMuon}}

The largest tensions in the current PDF fit involve the four muon-initiated 
fixed-target experiments ($\mu p \to \mu X$ and $\mu d \to \mu X$ measured 
by both BCDMS and NMC), as noted in Sec.\ \ref{sec:MuonExpts}.  
The kinematic regions covered by $\mu p \to \mu X$ and $e p \to e X$ 
experiments are shown in Fig.\ \ref{fig:figSix}.  
There is considerable overlap between the BCDMS and NMC experimental regions; 
but BCDMS extends farther toward $x = 1\,$, while NMC extends farther toward 
small $x$ and small $Q$.  Hence it is possible that the four muon experiments 
each measure different quantities.  Meanwhile, the H1 and ZEUS regions overlap 
completely with each other, and hardly at all with the muon experiments.  

\begingroup
\begin{table*}[htb]
  \begin{center}
\renewcommand\arraystretch{0.90}
\null
\begin{tabular}{||c|r|r|l||}
\hline
Expt & N & $\chi^2$ &  $(\gamma_1,\, \sigma_1), \, (\gamma_2,\, \sigma_2),\, \dots$ \\
\hline
BCDMS $F_2$p & 
339 & 384 & 
$(0.88,\,2.37)$ $(0.77,\,0.31)$ $(0.57,\,\mathbf{3.03})$ $(0.44,\,\mathbf{3.54})$ $(0.12,\,\mathbf{5.79})$ \\ 
BCDMS $F_2$d & 
251 & 248 & 
$(0.44,\,1.98)$ $(0.31,\,0.02)$ $(0.28,\,\mathbf{3.25})$ $(0.20,\,0.07)$ $(0.18,\,0.85)$ \\ 
NMC $F_2$p &  
201 & 332 & 
$(0.43,\,2.16)$ $(0.26,\,0.68)$ $(0.21,\,\mathbf{5.95})$ $(0.11,\,2.94)$ \\ 
NMC $F_2$p/d &  
123 & 121 & 
$(0.83,\,2.73)$ $(0.78,\,1.76)$ $(0.77,\,2.04)$ $(0.62,\,1.80)$ $(0.44,\,0.36)$ $(0.26,\,0.61)$ \\ 
\hline
\hline
BCDMS $F_2$p & 
339 & 365 & 
$(0.89,\,1.42)$ $(0.80,\,0.71)$ $(0.78,\,0.11)$ $(0.61,\,2.95)$ $(0.41,\,0.02)$ $(0.28,\,\mathbf{3.58})$ \\ 
BCDMS $F_2$d & 
251 & 249 & 
$(0.46,\,2.02)$ $(0.42,\,\mathbf{3.80})$ $(0.37,\,1.49)$ $(0.29,\,0.31)$ $(0.21,\,1.24)$ $(0.17,\,0.19)$ \\ 
NMC $F_2$p &  
201 & 331 & 
$(0.87,\,1.74)$ $(0.57,\,2.29)$ $(0.22,\,\mathbf{4.92})$ $(0.16,\,\mathbf{4.20})$ \\ 
NMC $F_2$p/d &  
123 & 118 & 
$(0.89,\,\mathbf{3.65})$ $(0.80,\,2.74)$ $(0.58,\,2.52)$ $(0.48,\,0.18)$ $(0.17,\,0.31)$ \\ 
\hline
BCDMS $F_2$p & 
339 & 365 & 
$(0.68,\,1.50)$ $(0.64,\,0.98)$ $(0.48,\,0.11)$ $(0.35,\,2.32)$ $(0.17,\,\mathbf{3.48})$ $(0.10,\,1.37)$ \\ 
BCDMS $F_2$d & 
251 & 260 & 
$(0.32,\,1.55)$ $(0.28,\,0.80)$ $(0.21,\,0.81)$ $(0.18,\,\mathbf{3.47})$ $(0.15,\,1.09)$ \\ 
NMC $F_2$p &  
201 & 338 & 
$(0.53,\,0.37)$ $(0.21,\,\mathbf{5.48})$ $(0.15,\,\mathbf{4.54})$ \\ 
NMC $F_2$p/d &  
123 & 119 & 
$(0.66,\,0.88)$ $(0.55,\,\mathbf{4.05})$ $(0.45,\,0.98)$ $(0.31,\,0.53)$ $(0.15,\,0.11)$ \\ 
\hline
$\;$ BCDMS $F_2$p & 
250 & 234 & 
$(0.69,\,0.42)$ $(0.60,\,0.99)$ $(0.46,\,0.24)$ $(0.31,\,1.11)$ $(0.12,\,2.04) \;$ \\ 
BCDMS $F_2$d & 
210 & 188 & 
$(0.29,\,1.64)$ $(0.26,\,2.25)$ $(0.20,\,0.91)$ $(0.18,\,2.16)$ $(0.16,\,2.79)$ $(0.12,\,2.29)$ \\ 
NMC $F_2$p &  
91 & 135 & 
$(0.20,\,2.01)$ $(0.11,\,1.71)$ \\ 
NMC $F_2$p/d &  
71 & 64 & 
$(0.59,\,2.59)$ $(0.45,\,1.87)$ $(0.32,\,0.78)$ $(0.22,\,1.11)$ \\ 
\hline
\end{tabular}
\vskip -10pt
  \end{center}
  \caption{Results with $\gamma_i > 0.1\,$ for the $\mu p$ and $\mu d$ 
experiments. In the first two groups, only one muon 
experiment---the one listed---is included in the global fit.  
In the last two groups, all four are included.
The first group uses the same parametrizations as Table \ref{table:table1}, 
while the other three groups use a parametrization with additional freedom 
for $u_v$ and $d_v$ at large $x$.  The fourth group includes the additional 
kinematic cuts shown in Fig.\ \ref{fig:figSix}.
  \label{table:table3}}
\vskip  5pt
\end{table*}
\endgroup


The observed tension involving the muon 
experiments could arise from inconsistencies within each muon 
data set, or disagreements between them, or disagreements between them and the
non-muon experiments.  The DSD method is an excellent tool to sort this out.

The first four lines of Table \ref{table:table3} show results from applying 
the DSD method to four new global fits, in which the experiment listed is the 
only one of the muon experiments included in the fit.  We see that large 
discrepancies---signaled by large $\sigma_i$---remain for the two $\mu p$ 
experiments.  Those discrepancies are further indicated by elevated values 
of $\chi^2/N$: $384/339$ and $332/201$.  This analysis removes the effect 
of any possible tension between the various muon experiments, so the 
discrepancies must be internal to each $\mu p$ data set, or else they 
reflect a conflict with the non-muon data.

Tension can also be created by insufficient flexibility in the 
functional forms that are used to approximate the PDFs at QCD scale $\mu_0$.  
In order to investigate this possible ``parametrization dependence,'' new fits 
were carried out in which two additional free parameters were introduced. The 
new parameters were added in 
$u_v(x) \equiv u(x)-\bar{u}(x)$ and $d_v(x) \equiv d(x)-\bar{d}(x)$, 
since these valence quark distributions dominate at large $x$, where the muon 
experiments are important according to Fig.\ \ref{fig:figSix}.  The second 
group of four lines in Table \ref{table:table3} shows the DSD results for these 
fits, where again only one muon experiment is included in each fit.  
The additional freedom produces a better agreement with the BCDMS $\mu p$ 
experiment, as $\chi^2$ drops from 384 to 365 for 339 data points.
However, the $\sigma_i$ show that substantial tensions remain for all four of 
the muon experiments.

The third group of four lines in Table \ref{table:table3} shows results from 
a fit that once again includes all four muon experiments, as in 
Table \ref{table:table1}; but includes the two new valence-quark parameters.  
One sees that this more flexible parametrization does not eliminate the 
tension.  Further improvement cannot be obtained by further increasing the 
flexibility of the parametrization, because attempts to do that are foiled 
by the fits becoming unstable, due to large undetermined parameters.

It was possible to shed further light on the source of tension in the muon data 
sets by splitting each set into a low-Q and high-Q region.  When this was done 
(not shown), it was found that both the low-Q and the high-Q portions of each 
muon experiment are separately consistent with the non-muon data.  
\textit{Hence the observed tension is generated by the Q-dependence of 
each muon data set.}  

This is actually very plausible, because the BCDMS data have previously been 
shown \cite{HigherTwist} to contain a significant ``higher-twist'' component 
(non-leading power-law dependence in $Q$), which is not taken into account in 
the PDF fit.  Higher-twist effects can be expected to be even more important 
for the NMC data, since more of that data is at small $Q$.

To suppress higher-twist contributions---or other possible deviations from 
NLO QCD at low $Q$---we now remove the BCDMS data 
that lie below the dashed line and the NMC data that lie below the solid 
line in Fig.\ \ref{fig:figSix}.  (These cuts were chosen roughly
based on \cite{HigherTwist}; they have not been optimized.) 
The resulting fit, which includes all four of the muon data sets, is summarized 
by the final group of four lines in Table \ref{table:table3}.  With the cuts, 
the large tension has gone away.  
The $\chi^2/N$ for the BCDMS experiments is also greatly improved, and is now 
within the normal range.  The $\chi^2/N$ is still a bit high for the NMC 
$\mu p$ data, which suggests that a somewhat stronger cut would be desirable 
for that data set. 


\begin{figure*}[tbh]
\vskip 10pt
\hfill
 \resizebox*{0.45\textwidth}{!}{
\includegraphics[clip=true,scale=1.0]{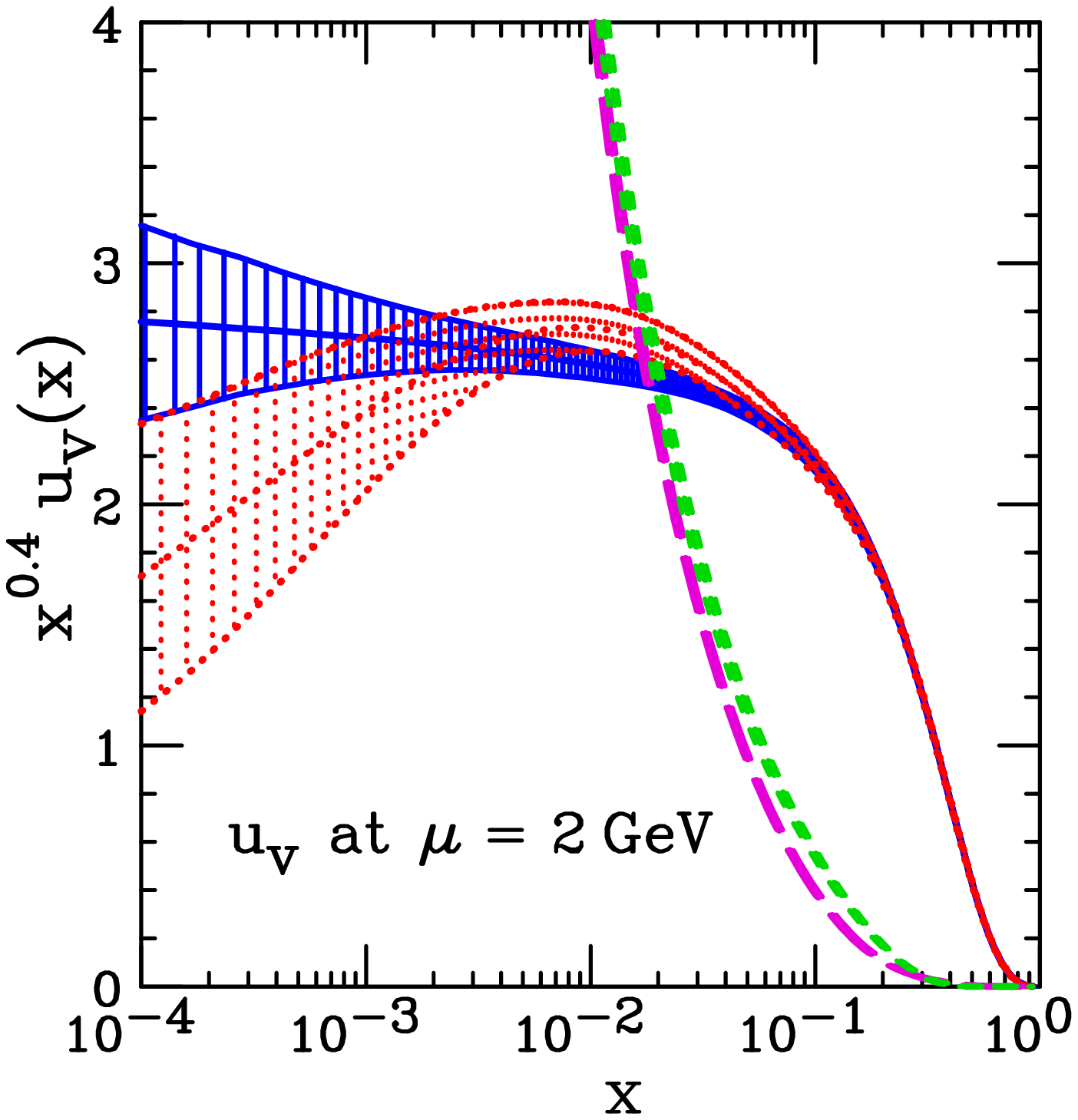}}
\hfill
 \resizebox*{0.45\textwidth}{!}{
\includegraphics[clip=true,scale=1.0]{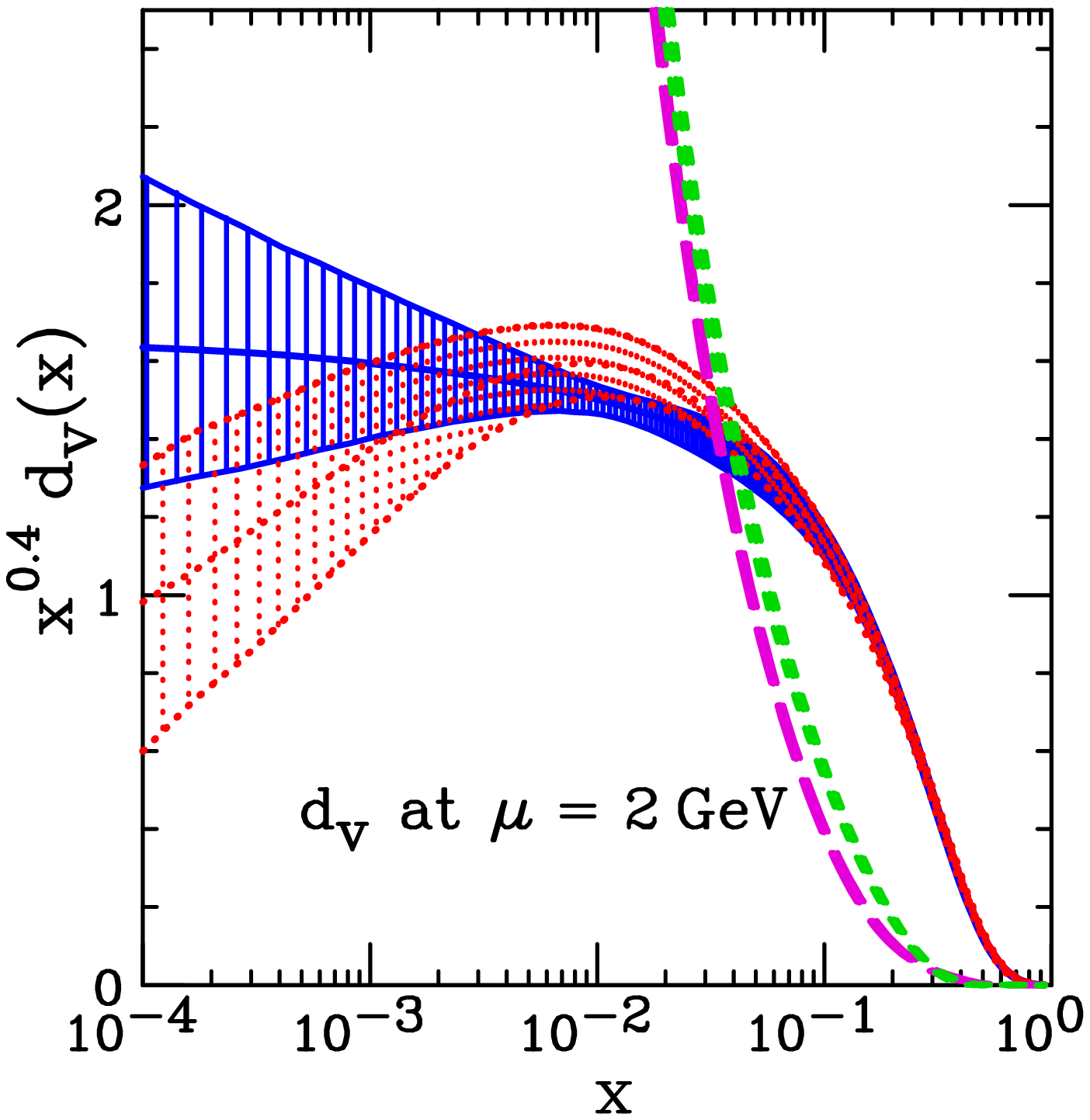}}
\hfill
\vskip -5pt
 \caption{Uncertainty of valence quark distributions at $\Delta\chi^2 = 10$. 
The solid shaded region is the fit in Table \ref{table:table1}. The dotted 
region is the fit in the last section of Table \ref{table:table3}.  
The long-dashed and short-dashed curves show $\bar{u}$ and $\bar{d}$ 
respectively for comparison.
}
 \label{fig:figSeven}
\vskip  5pt
\end{figure*}

\section{Implications for PDF analysis
\label{sec:implications}}

The results presented here support a tolerance criterion of 
$\Delta\chi^2 \approx 10$ to estimate the 90\% confidence range of PDF 
uncertainties, based solely on the uncertainties of the input data.  In 
contrast, PDF determinations made using the Hessian method \cite{Hessian} 
generally include a much broader allowed range of uncertainty, 
\textit{e.g.}\
$\Delta\chi^2 = 50$ in MRST \cite{MRST2004} and 
$\Delta\chi^2 = 100$ for 90\% confidence in CTEQ \cite{cteq66}.  
This larger range arises from 
adopting a ``hypothesis-testing'' criterion \cite{Collins}, according to which 
any PDF configuration that provides a satisfactory fit to all of the
input data sets is deemed acceptable.  Loosely speaking, the hypothesis-testing 
criterion is defined by $\chi^2 \, < \, N + \sqrt{2 N}$ for each experiment. 
The overall allowed $\Delta\chi^2$ is therefore 
$\approx \sqrt{2 N_{\mathrm{total}}}$, \textit{i.e.,} 77 for 3000 data 
points at $1 \, \sigma \,$.

In detail, the hypothesis-testing condition is corrected for finite $N$ for
each experiment, and 
refined on the basis of the lowest possible $\chi^2$ that can be achieved for 
that experiment.  In the CTEQ fits \cite{cteq66}, contributions 
to $\chi^2$ from some of the data sets are enhanced by weight factors, which 
are chosen to keep the fits to those experiments adequate over the 
$\chi_{\mathrm{tot}}^{\,2} \, < \, \chi_{\mathrm{min}}^{\, 2} \, + \, 
\Delta\chi^2$ range.
In the most recent CTEQ fit \cite{CT09}, that procedure is supplemented by 
adding a quartic penalty term to the effective $\chi^2$, to force the fits to 
some recalcitrant experiments to remain satisfactory over the entire region 
defined by $\Delta\chi^2$.  Meanwhile, recent MSTW 
fits \cite{MSTW08,MSTWalphas} abandon the use of a fixed $\Delta\chi^2$, and 
instead determine the uncertainty limit ``dynamically'' along each eigenvector 
direction (separately for ``$+$'' and ``$-$'' senses), as the point where the 
fit first becomes unacceptable to one of the data sets.

The hypothesis-testing criterion is a minimal requirement for acceptable fits.  
It defines a broader uncertainty limit than would be predicted on normal
statistical grounds, which has been called the ``parameter-fitting'' 
criterion \cite{Collins}.  The parameter-fitting criterion 
is ideally defined by $\Delta\chi^2 = 1.0$ ($2.7$) for a 68\% (90\%) confidence 
interval.  In view of the results of this paper, that should be 
expanded in practice to $\Delta\chi^2 \approx 10$ for 90\% confidence, 
on the basis of inconsistencies observed among the implications 
of different data sets.

From a statistical point of view, the hypothesis-testing criterion appears to 
be overly conservative.  That notion is challenged, however, by the apparent 
``time-dependence'' and ``space-dependence'' of the PDFs.  Namely, we have 
repeatedly seen changes from one generation of PDFs to the next, \textit{e.g.}, 
CTEQ5/\-CTEQ6.0/\-CTEQ6.1/\-CTEQ6.6/\-CT09 or
MRST2001/\-MRST2002/\-MRST2004/\-MSTW2008,
for which the central estimate for 
some flavor in a set is close to the predicted 90\% confidence limit from the 
previous set; and differences between PDF sets determined by different groups, 
such as CTEQ and MSTW, are also frequently as large as these broad uncertainty 
estimates.  Examples of this can be seen in Figs.\ 8 and 12 of \cite{CT09}.

Some of the time-dependence has resulted from improvements in the theory, 
such as the better treatment of heavy quark mass effects beginning with 
CTEQ6.6; or additions to the available data.  But other differences between 
PDF determinations arise from the 
choices of which data sets to include; in choices of kinematic cuts such 
as those introduced in Sec.\ \ref{sec:MoreMuon} to remove data points for 
which the perturbative QCD treatment is suspect; and in the choice of 
parametrizing functions at $\mu_0$.  Additional uncertainties are present 
due to the NLO approximations made in the theory.  Adopting the 
hypothesis-testing criterion can be seen as an expedient way to broaden the 
estimated uncertainty range to allow for these uncertainties---although 
obviously they cannot be reliably predicted on the basis of the errors in 
the experimental data sets. 

Quantities that are weakly constrained by the data are especially subject to 
parametrization dependence.  A classic example of this is provided by the 
gluon distribution at large $x$.  Prior to measurements of the inclusive jet 
cross section at the Tevatron, there was very little information on the gluon 
at large $x$.  The parametrizations used at that time therefore devoted very 
few parameters to the large-$x$ region, since unconstrained parameters make 
the fitting procedure unstable.  When the jet data became available, they were 
found to lie outside the predicted uncertainty range.  This pointed to a need 
to introduce additional parameters, which could then be determined by stable 
fits using the new data.

A further example is shown in Fig.\ \ref{fig:figSeven}, which displays the 
valence quark distributions 
$u_v \equiv u - \bar{u}$ and 
$d_v \equiv d - \bar{d}$.
Their uncertainty is large for $x \, \lesssim \, 0.01$, because their 
contribution from that region to most observables is swamped by much larger 
contributions from $\bar{u}$ and $\bar{d}$, which are also shown in the 
figure.  The uncertainties are shown for both the original fit (described 
in Table \ref{table:table1}) and the final fit (described in the last four 
lines of Table \ref{table:table3}).  The final central fit lies far outside 
the uncertainty band estimated in the earlier fit, when that uncertainty is 
computed at $\Delta\chi^2 = 10$.  At large $x$, the situation is 
reversed: valence quarks dominate the phenomenology, so $u_v$ and $d_v$ are 
very well measured there, and the difference between the two fits is small 
and consistent with the estimated uncertainty.  

An emerging alternative to the Hessian approach is provided by the 
NNPDF \cite{NNPDF} method, which avoids the parametrization problem by using 
very flexible Neural Network representations instead of functional forms to 
describe the PDFs at $\mu_0$.  
An attractive feature of the NNPDF approach is that introducing new 
measurements reduces the uncertainty of the output PDFs, unless the new data 
are rather inconsistent with the previous data.  The same cannot be said for 
the Hessian approach, because when new data sets are added to the global fit, 
it is often desirable to increase the flexibility of the parametrization, as 
happened with the inclusive jet cross section as discussed above.  The NNPDF 
method incorporates experimental errors by creating an ensemble of fits to 
``pseudodata'' sets in which the measured values are displaced by random 
shifts that are proportional to the experimental uncertainties.  It would be 
interesting to apply the NNPDF method to assess the uncertainties that are 
\textit{not} associated with the experimental errors, by using the original 
unshifted data to produce each element in the ensemble.  The ensemble would 
then retain the other sources of uncertainty due to the other random processes 
used to create it.

The effective number of parton parameters that can be measured by the available 
data---currently around 25---is small enough that the traditional Hessian 
method is convenient.  But the number of potential parameters that could be 
determined by some future experiment, but which are currently unconstrained, 
is of course very large or even infinite.  So it is not practical to provide 
parameters for all such potential degrees of freedom.  However, the Hessian 
approach may nevertheless be viable for the large range of predictions for 
which PDFs are needed, since the processes one wishes to predict depend on 
similar aspects of the PDFs to the experiments that are used to determine 
them.  As an extreme case in point, the PDF fits described in this paper admit 
no uncertainty at all in the assumption $s(x,\mu) = \bar{s}(x,\mu)$.  They can
therefore not be used to predict new processes that are sensitive to the 
strangeness asymmetry $s^{(-)}(x,\mu) \equiv s(x,\mu) - \bar{s}(x,\mu)$; but 
most processes we wish to predict are not in fact sensitive to that asymmetry.


\section{Conclusion 
\label{sec:conclusion}}

The recently-developed DSD method \cite{DSD} has been applied to assess 
compatibility among the data sets that are used to extract parton distribution 
functions.  The DSD method is more discerning than the previous 
method \cite{Collins,CT09} of studying correlations between $\chi^2$ values 
for the various experiments, because it looks for inconsistencies of each 
experiment along the specific directions in parameter space for which that 
experiment is significant in the global fit, while ignoring the large number 
of directions along which the experiment is unimportant.

Results from the DSD method, which are shown in Table \ref{table:table1}, 
can be read as a ``report card''  on the contribution of each experiment to 
the global fit.  The $\gamma_i$ parameters 
measure how much each experiment influences the fit, while the $\sigma_i\,$ 
parameters measure how much dissonance each experiment brings with it. 

Table \ref{table:table1} identified fixed-target $\mu p$ and $\mu d$ 
experiments as the greatest source of tension in a recent global fit.  
Further exploration in Sec.\ \ref{sec:MoreMuon} revealed the underlying cause 
of that tension as deviations from NLO QCD predictions---presumably due to 
higher twist---which had previously been observed in these 
data \cite{HigherTwist}, but which were not taken into account in the 
fit.  Kinematic cuts shown in Fig.\ \ref{fig:figSix} 
remove the contaminated region and eliminate 
the large discrepancies, as can be seen by comparing the last four lines of 
Table \ref{table:table3} with their corresponding entries in 
Table \ref{table:table1}.  Future global fits should make a refined version 
of these cuts, or else introduce additional fitting parameters to model the 
higher-twist contribution.  (The latter was attempted in CTEQ6 \cite{cteq60}, 
without conclusive results---the DSD method not being available at that time 
to make a sensitive test of the consistency.)

Independently of the muon experiments, the implications of the various data 
sets in the global fit are found to be somewhat inconsistent with each other 
(Fig.\ \ref{fig:figFive}). The average discrepancy is a bit less than a factor 
of 2 larger than what is predicted by straightforward propagation of the 
experimental errors.  This was shown in Sec.\ \ref{sec:NonGaussian} to suggest 
that the 90\% confidence limit for predictions from the global fit should be 
estimated by a tolerance criterion of $\Delta\chi^2 \approx 10$, in place of 
the $\Delta\chi^2 = 2.71$ that would be implied by pure Gaussian statistics.

Much larger tolerance criteria ($\Delta\chi^2 = 50$ \cite{MRST2004} or 
$\Delta\chi^2 = 100$ \cite{cteq60,cteq66}) have been used to estimate the 90\% 
confidence limit in recent applications of the Hessian approach.  These 
more conservative tolerance criteria correspond to the ``hypothesis testing''
notion that any PDF set is acceptable as long as its fit to every data set lies
in the nominal statistical range $\chi^2 = N \, \pm \, \sqrt{2N}$, or its 
90\% analog, with appropriate corrections for finite $N$. This implies 
an effective overall 
$\Delta\chi_{\mathrm{total}}^2 \sim \sqrt{2 N_{\mathrm{total}}} \approx 75$ 
for $1 \, \sigma$.  
The uncertainty from input data, as assessed in this paper by studying its 
mutual consistency, does not call for this expanded uncertainty range.  
However, some aspects of the fit \textit{do} no doubt require an expanded 
uncertainty estimate, because of theoretical systematic errors---most notably 
the use of NLO perturbation theory and parametrization dependence.  
The need for some such expanded uncertainty is demonstrated by the relatively 
large changes in uncertainty bands that can be caused by relatively minor 
changes in the choice of parametrization or in the choice of data sets that 
are included.  Examples of this are provided by the valence quark 
distributions at small $x$, as discussed in Sec.\ \ref{sec:MoreMuon}; and the 
gluon distribution, as discussed in \cite{CT09}.


In the future, it would be desirable to estimate the uncertainties associated 
with parametrization choices and other theoretical errors directly, rather 
than using a large $\Delta\chi^2$ to stand in for them in a manner that is 
based artificially on the uncertainties of the data.  If this can be 
accomplished, the result will likely expand the estimated uncertainty range 
for quantities that are poorly constrained; but it may reduce the uncertainty 
for quantities that are well constrained, because of the reduction in 
$\Delta\chi^2$.  From the ratio of $\Delta\chi^2$ values, one might hope to 
find the uncertainty reduced by as much as a factor of 3; but the actual 
reduction will probably be less than that, because $\chi^2$ generally rises 
faster than quadratic for large displacements from the best fit.




\begin{acknowledgments}
I thank my 
TEA (Tung et al.)
colleagues J.\ Huston, H.~L.~Lai, P.\ M.\ Nadolsky, and C.-P.~Yuan
for discussions of these issues.
I thank Louis Lyons for discussions and for suggesting the illustrative 
example that is described in the Appendix.
This research was supported by National Science Foundation grant PHY-0354838. 
\end{acknowledgments}

\section*{Appendix}
A formal derivation of the DSD method was presented in \cite{DSD}, and it is 
reviewed in Sec.\ \ref{sec:DSDmethod}.  To assist in understanding the method, 
this Appendix illustrates it by a simple explicit example.  

Suppose we have a theory that predicts a linear relationship $y = ax + b$ 
where $a$ and $b$ are unknown parameters.  Further suppose there are three 
experiments, which have measured
\begin{eqnarray}
  \mbox{Expt 1:}\; \; y &=& y_1 \, \pm 1 \; \mbox{ at } x = -1 \nonumber \\
  \mbox{Expt 2:}\; \; y &=& y_2 \, \pm 1 \; \mbox{ at } x = \; \; 0 \nonumber \\
  \mbox{Expt 3:}\; \; y &=& y_3 \, \pm 1 \; \mbox{ at } x = \; \; 1  \; .
  \label{eq:app1}
\end{eqnarray}
The fit to these three experiments is described by 
\begin{equation}
\chi^2 = \chi_1^{\, 2} \, + \, \chi_2^{\, 2} \, + \, \chi_3^{\, 2} \;,
  \label{eq:app2}
\end{equation}
where
\begin{eqnarray}
  \chi_1^{\, 2} &=& \left(\frac{y_1 \, - \, (-a+b)}{1}\right)^2 \nonumber \\
  \chi_2^{\, 2} &=& \left(\frac{y_2 \, - \, (b)}{1}\right)^2 \nonumber \\
  \chi_3^{\, 2} &=& \left(\frac{y_3 \, - \, (a+b)}{1}\right)^2 \; .
  \label{eq:app3}
\end{eqnarray}
It is natural to replace the theory parameters $a$ and $b$ by new parameters 
$u_1$ and $u_2$ that are measured from the minimum point in $\chi^2$ and 
normalized in the standard Hessian way:
\begin{eqnarray}
  a &=& \frac{u_1}{\sqrt{2}} \, + \, \frac{y_3 - y_1}{2} \nonumber \\
  b &=& \frac{u_2}{\sqrt{3}} \, + \, \frac{y_1 + y_2 + y_3}{3} \; .
  \label{eq:app4}
\end{eqnarray}
This puts $\chi^2$ into the standard diagonal form
\begin{equation}
  \chi^2  \, = \, u_1^{\,2} \, + \, u_2^{\,2} \, + \, K^2 \; ,
  \label{eq:app5}
\end{equation}
where 
\begin{equation}
K \, = \, (y_1 + y_3 - 2y_2)/\sqrt{6} \;.
  \label{eq:app6}
\end{equation}
The transformation (\ref{eq:app4}) that yields (\ref{eq:app5}) contains 
a shift and a rescaling of the original fitting parameters $a$ and $b$.  
In general, it also requires a rotation (orthogonal transformation) that 
intermingles those variables; but that was not necessary in this simple 
example because of symmetry.  The uncertainty on the theory parameters 
$a$ and $b$ in the global fit can now be obtained easily from 
Eq.\ (\ref{eq:app5}), which implies that the $1 \, \sigma$ limits are 
$u_1 = 0 \pm 1$ and $u_2 = 0 \pm 1$.

In order to examine the internal consistency of this fit, we must consider the 
contributions to $\chi^2$ from the individual experiments. These can be 
expressed in terms of the new coordinates as
\begin{eqnarray}
  \chi_1^{\,2} &=& \left(\frac{u_1}{\sqrt{2}} \, - \, \frac{u_2}{\sqrt{3}} 
  \, + \,\frac{K}{\sqrt{6}}\right)^2 \nonumber \\
  \chi_2^{\,2} &=& \left(\frac{u_2}{\sqrt{3}} \, + \, \frac{2 K}{\sqrt{6}}\right)^2 \nonumber \\
  \chi_3^{\,2} &=& \left(\frac{u_1}{\sqrt{2}} \, + \, \frac{u_2}{\sqrt{3}} 
        \, - \, \frac{K}{\sqrt{6}}\right)^2 \; .
  \label{eq:app7}
\end{eqnarray}
To study the consistency between Expt 1 and its complement, it is necessary 
to make a further coordinate transformation to diagonalize $\chi_1^{\,2}\,$.  
That transformation can be found by the DSD method; or in this simple case, 
by inspection:
\begin{eqnarray}
  u_1 &=&  \quad \sqrt{3/5} \, v_1 \, + \, \sqrt{2/5} \, v_2  \nonumber \\
  u_2 &=& -\,\sqrt{2/5} \, v_1 \, + \, \sqrt{3/5} \, v_2 \; .
  \label{eq:app8}
\end{eqnarray}
This gives
\begin{eqnarray}
  \chi_1^{\,2} &=& \left(\frac{v_1 \, + \, \sqrt{1/5} \, K}{\sqrt{6/5}}\right)^2 \nonumber \\
  \chi_{\overline{1}}^{\,2} \, = \, 
  \chi_2^{\,2} \, + \, \chi_3^{\,2} &=& 
 \left(\frac{v_1 \, - \, \sqrt{5} \, K}{\sqrt{6}}\right)^2 \, + \, 
           v_2^{\,2} \nonumber \\
  \chi^2 = \chi_1^{\,2} \, + \, \chi_{\overline{1}}^{\,2} &=& 
           v_1^{\,2} \, + \, v_2^{\,2} \, + K^2 \;.
  \label{eq:app9}
\end{eqnarray}
From this, one easily reads
\begin{eqnarray}
  \mbox{Expt 1:}\quad v_1  &=& -\,\sqrt{1/5} \, K  \, \pm  \, \sqrt{6/5} \nonumber \\
  \mbox{Expt $\overline{\mbox{1}}$:}\quad v_1 &=& +\,\sqrt{5} \, K \, \pm  \, \sqrt{6} \; .
  \label{eq:app10}
\end{eqnarray}
Subtracting these two measurements and combining their errors in quadrature 
shows that they differ by $\sqrt{36/5} \, K  \, \pm  \, \sqrt{36/5}$.  This 
differs from $0$ by $K\,$ standard deviations, which is the measure of 
consistency between Expt 1 and its complement.

The contribution to $\chi^2$ from Expt 2 in Eq.\ (\ref{eq:app7}) happens to be 
already in the diagonal form that is the heart of the DSD method, so it 
requires no further transformation: 
\begin{eqnarray}
  \chi_2^{\,2} &=& \left(\frac{u_2 \, + \, \sqrt{2}K}{\sqrt{3}}\right)^2 \nonumber \\
  \chi_{\overline{2}}^{\,2} \, = \, \chi_1^{\,2} \, + \, \chi_3^{\,2} &=& u_1^{\,2} \, + \, 
          \left(\frac{u_2 \, - \, \sqrt{1/2}\,K}{\sqrt{3/2}}\right)^2 \; .
  \label{eq:app11}
\end{eqnarray}
From this, one reads
\begin{eqnarray}
  \mbox{Expt 2:}\quad u_2  &=& -\,\sqrt{2}\, K  \, \pm \, \sqrt{3} \nonumber \\
  \mbox{Expt $\overline{\mbox{2}}$:}\quad u_2 &=& +\,\sqrt{1/2}\, K \, \pm \, \sqrt{3/2} \; .
  \label{eq:app12}
\end{eqnarray}
Subtracting these results and combining their errors in quadrature shows that 
the measurement of $u_2$ by Expt 2 and its complement differ by 
$\sqrt{9/2}\,K \, \pm \, \sqrt{9/2}$.  This difference is also $K$ standard 
deviations away from $0\,$.

Because there are only three data points in this example, with two free 
parameters in the theory, there is only one possible test of the internal 
consistency.  That is why both Expt 1 and Expt 2 show the same discrepancy $K$,
when the discrepancy is measured in standard deviations.  To show that Expt 3 
would also give the same result is left as an exercise for the reader! 

In this simple example, the consistency measure can also be found by elementary 
means:  adding the errors from (\ref{eq:app1}) in quadrature gives 
$\pm \, \sqrt{6}$ for the uncertainty of $y_1 + y_3 - 2 y_2$, and hence the 
uncertainty of $K$ is $\pm 1$ by Eq.\ (\ref{eq:app6}).  Meanwhile, the 
theoretical prediction for $K$ is $0$, since the theory predicts $y$ to be a 
linear function of $x$, and $y_1$, $y_2$, $y_3$ are measured symmetrically at 
$x = -1, 0, 1$.  Hence the difference between theory and experiment is 
$K \pm 1$, so $K$ is indeed the discrepancy measured in standard deviations.

\end{document}